\documentclass{eptcs}

\usepackage[table]{xcolor}

\usepackage{amsbsy}
\usepackage{amsmath}
\usepackage{amsfonts}
\usepackage{amssymb}
\usepackage{amsthm}
\usepackage{bm}
\usepackage{dashrule}
\usepackage{reo-tikz}
\usepackage{xspace}

\providecommand{\event}{ICE 2011} 

\newtheorem{definition}{Definition}
\newtheorem{lemma}{Lemma}

\newtheorem{proposition}{Proposition}

\newcommand{\pcomment}[2][]	{\color{gray} \mbox{\begin{minipage}[t]{.975\textwidth}
								\vspace*{-.5em}\sf
								$/^*$ #2 \rotatebox[origin=c]{180}{$/^*$} 
								\vspace*{.25em} 
							\end{minipage}}}

\newcommand{\barproof}[2][]	{%
	\begin{tabular}{@{\hspace{-.9em}}cl@{}}	
		\cellcolor{lightgray}\hspace{-.9em}
		&
		\begin{minipage}{.97\textwidth}\begin{proof}#2\end{proof}\end{minipage}
	\end{tabular}
}

\newcommand{\setbar}				{\hspace{.1cm}|\hspace{.1cm}}
\newcommand{\setleft}				{\{\hspace{.05cm}}
\newcommand{\setright}				{\hspace{.05cm}\}}

\newcommand{\tuple}[2][]			{\langle #2\rangle}
\newcommand{\tablefill}[2][]		{\begin{minipage}[t]{0pt}\vspace*{#2}\end{minipage}}

\newcommand{\TITLE}{Correlating Formal Semantic Models of Reo Connectors: Connector Coloring and Constraint Automata}
\title{\TITLE}
\author{%
	Sung-Shik T.Q. Jongmans
	\institute{Centrum Wiskunde \& Informatica (CWI) \\ Amsterdam, the Netherlands}
	\email{jongmans@cwi.nl}
	\and
	Farhad Arbab
	\institute{Centrum Wiskunde \& Informatica (CWI) \\ Amsterdam, the Netherlands}
	\email{farhad.arbab@cwi.nl}
}

\def\titlerunning{Correlating Formal Semantic Models of Reo Connectors}
\def\authorrunning{S.-S.T.Q. Jongmans \& F. Arbab}

\begin{document}

\maketitle

\begin{abstract}
Over the past decades, coordination languages have emerged for the specification and implementation of interaction protocols for communicating software components. This class of languages includes Reo, a platform for compositional construction of connectors. In recent years, various formalisms for describing the behavior of Reo connectors have come to existence, each of them serving its own purpose. Naturally, questions about how these models relate to each other arise. From a theoretical point of view, answers to these questions provide us with better insight into the fundamentals of Reo, while from a more practical perspective, these answers broaden the applicability of Reo's development tools. In this paper, we address one of these questions: we investigate the equivalence between coloring models and constraint automata, the two most dominant and practically relevant semantic models of Reo. More specifically, we define operators that transform one model to the other (and vice versa), prove their correctness, and show that they distribute over composition. To ensure that the transformation operators map one-to-one (instead of many-to-one), we extend coloring models with data constraints. Though primarily a theoretical contribution, we sketch some potential applications of our results: the broadening of the applicability of existing tools for connector verification and animation.
\end{abstract}

\raggedbottom

%
\section{Introduction}
\label{sect:intr}

Over the past decades, \emph{coordination languages} have emerged for the specification and implementation of interaction protocols for communicating software components. This class of languages includes Reo \cite{arbab04}, a platform for compositional construction of \emph{connectors}. Connectors in Reo  (or \emph{circuits}) form the communication mediums through which components can interact with each other. Essentially, Reo circuits impose constraints on the order in which components can send and receive \emph{data items} to and from each other. Although ostensibly simple, Reo connectors can describe complex protocols (e.g., a solution to the Dining Philosophers problem \cite{arbab05}). In recent years, various formal models for describing the behavior of Reo circuits have arisen, including a coalgebraic model \cite{arbab03}, various operational models (e.g., \emph{constraint automata} \cite{baier06}), and two coloring models \cite{clarke07} (we mention more models in Section \ref{sect:conc}). Each of these formalisms serves its own purpose: the coalgebraic model has become Reo's reference semantics, constraint automata play a dominant role in connector verification (e.g., the Vereofy model checker \cite{baier09}), and the coloring models facilitate the animation of connectors (e.g., the implementation of the work in Chapter 6 of \cite{costa10} in the Eclipse Coordination Tools).

Having this wide variety of semantic models, questions about how they relate to each other naturally arise. We identify two reasons for why this question, moreover, requires answering. First, from a purely theoretical point of view, answers provide us with better and possibly new insights into Reo's fundamentals. Second, from a more practical perspective, such answers broaden the applicability of tools|both existing and future|that assist developers in designing their Reo circuits. For instance, the correspondence between constraint automata and the coloring model with two colors (the topic of this paper) enables us to, on the one hand, model check connectors with a coloring model as their formal semantics, and, on the other, animate connectors with a constraint automaton as their behavioral model. We consider this important because contemporary tools for verification and animation cannot operate on, respectively, coloring models and constraint automata.%

\paragraph{Contributions}

We investigate the relation between coloring models with two colors and constraint automata. We show how to transform the former to the latter and demonstrate bi-similarity between an original and its transformation. In the opposite direction, we show how to transform constraint automata to equivalent coloring models, prove that these transformations define each other's inverse, and again show bi-similarity. Additionally, we prove the compositionality of our transformation operators. To ensure that our transformation operators map one-to-one (instead of many-to-one), we extend coloring models with data constraints. To illustrate the practical relevance of our work, we sketch one of its applications: the integration of verification and animation of context-sensitive connectors in Vereofy. We emphasize, however, that with this paper, we aim at establishing equivalences: we consider it primarily a theoretical contribution and a formal foundation for future tool development.%

\paragraph{Organization}

In Section \ref{sect:prelim}, we discuss preliminaries of Reo. In Section \ref{sect:constraintcc}, we extend coloring models with data constraints. In Section \ref{sect:cctoca}, we present a transformation from such data-aware coloring models to constraint automata, and in Section \ref{sect:catocc}, we present a transformation in the opposite direction. In Section \ref{sect:appl}, we sketch an application of our results. Section \ref{sect:conc} concludes the paper and includes related work.
\newcommand{\exclrouter}			{\textsf{ExclRouter}\xspace}
\newcommand{\fifo}					{\textsf{FIFO}\xspace}
\newcommand{\lossyfifo}				{\textsf{LossyFIFO}\xspace}
\newcommand{\lossysync}				{\textsf{LossySync}\xspace}
\newcommand{\merger}				{\textsf{Merger}\xspace}
\newcommand{\replicator}			{\textsf{Replicator}\xspace}
\newcommand{\sync}					{\textsf{Sync}\xspace}
\newcommand{\syncdrain}				{\textsf{SyncDrain}\xspace}
\newcommand{\syncfifo}				{\textsf{SyncFIFO}\xspace}

\newcommand{\exclrouterindex}		{\texttt{ExclRouter}\xspace}
\newcommand{\fifoindex}				{\texttt{FIFO}\xspace}
\newcommand{\fifoemptyindex}		{\texttt{FIFO-E}\xspace}
\newcommand{\fifofullindex}			{\texttt{FIFO-F}\xspace}
\newcommand{\lossyfifoindex}		{\texttt{LFIFO}\xspace}
\newcommand{\lossyfifoemptyindex}	{\texttt{LFIFO-E}\xspace}
\newcommand{\lossyfifofullindex}	{\texttt{LFIFO-F}\xspace}
\newcommand{\lossysyncindex}		{\texttt{LSync}\xspace}
\newcommand{\mergerindex}			{\texttt{Merger}\xspace}
\newcommand{\replicatorindex}		{\texttt{Repl}\xspace}
\newcommand{\syncindex}				{\texttt{Sync}\xspace}
\newcommand{\syncdrainindex}		{\texttt{SyncDrain}\xspace}
\newcommand{\syncfifoindex}			{\texttt{SyncFIFO}\xspace}

%
\section{Connector Structures, Coloring Models, and Constraint Automata}
\label{sect:prelim}

In this section, we discuss the essentials of Reo (relevant to this paper): the structure of circuits and two formal models of its behavior. Henceforth, we write ``connector'' or ``circuit'' to refer to both the structure and the intended behavior of a communication medium between software components.

\newcommand{\conn}							{\mathcal{C}}
\newcommand{\conncomp}						{\times}
\newcommand{\conncol}						{\mathcal{C}^{\mbox{\upshape\tiny\textsf{Col}}}}
\newcommand{\conncolcomp}					{\conncomp}
\newcommand{\connca}						{\mathcal{C}^{\mbox{\upshape\tiny\textsf{CA}}}}
\newcommand{\conncacomp}					{\conncomp}
\newcommand{\connstruct}					{\sigma}
\newcommand{\connstructcomp}				{\boxtimes}

\newcommand{\iomap}							{E}

\newcommand{\primstruct}					{e}
\newcommand{\primstructset}					{E}

\newcommand{\reonode}						{n}
\newcommand{\reonodeset}					{N}
\newcommand{\reobnodeset}					{B}
\newcommand{\reofnodeset}					{F}
\newcommand{\reonodeuniverse}				{\mbox{\upshape\textsf{Node}}}

\subsection{Connector Structures}
\label{sect:prelim:cs}

We start with the structure of circuits. A Reo connector consists of \emph{nodes} through which data items can \emph{flow}. We distinguish three types of nodes: \emph{input nodes} on which components can issue \emph{write requests} for data items, \emph{internal nodes} that the connector uses to internally route data items, and \emph{output nodes} on which components can issue \emph{take requests} for data items. We call input and output nodes collectively, the \emph{boundary nodes} of a connector. Write and take requests, collectively called \emph{I/O requests}, remain \emph{pending} on a boundary node until they \emph{succeed}, in which case their respective nodes \emph{fire}. We describe the structure of a connector formally as a set of nodes, typically denoted as $\reonodeset$, and a binary relation on these nodes, typically denoted as $\iomap$. We use this relation to specify the direction of the flow through the nodes in $\reonodeset$: if $\tuple{\reonode_{in} , \reonode_{out}} \in \iomap$, this means that node $\reonode_{in}$ can route incoming data items from itself to $\reonode_{out}$.%
\begin{definition}
	[Universe of nodes]
	$\reonodeuniverse$ is the set of nodes.
\end{definition}
\begin{definition}
	[Connector structure]
	\label{def:connstruct}
	A connector structure $\connstruct$ is a pair $\tuple{\reonodeset , \iomap}$ with $\reonodeset \subseteq \reonodeuniverse$ a set of nodes and $\iomap \subseteq \reonodeset \times \reonodeset$ a relation such that $\reonode \in \reonodeset$ implies $\tuple{\reonode_{in} , \reonode} \in \iomap \mbox{ or } \tuple{\reonode , \reonode_{out}} \in \iomap$.
\end{definition}
\noindent The side condition in the previous definition ensures that a connector structure does not include superfluous nodes through which data items never flow (note that it does allow for cyclic structures). We associate the following sets and definitions with a connector structure $\connstruct = \tuple{\reonodeset , \iomap}$. First, we define the sets of its input, output, and internal nodes.%
\begin{center}
	$\begin{array}{lcl}
		\texttt{input}_\connstruct & = & \setleft \reonode \in \reonodeset \setbar \tuple{\reonode_{in} , \reonode} \notin \iomap \setright
	\\	\texttt{output}_\connstruct & = & \setleft \reonode \in \reonodeset \setbar \tuple{\reonode , \reonode_{out}} \notin \iomap \setright
	\\	\texttt{internal}_\connstruct & = & \setleft \reonode \in \reonodeset \setbar \tuple{\reonode_{in} , \reonode} , \tuple{\reonode , \reonode_{out}} \in \iomap \setright
	\end{array}$
\end{center}
Note that these three definitions specify mutually disjoint sets and that their union equals $\reonodeset$ (due to the side condition in Definition \ref{def:connstruct}). If $\texttt{internal}_\connstruct = \emptyset$, the circuit whose topology $\connstruct$ describes belongs to the class of connectors called \emph{primitives}, the most elementary mediums between components. 

\begin{figure}[t]
	\centering
	\scalebox{.8}{
	\begin{tabular}{|c|c|cc|}
		\hline &&&
	\\	\vspace{-1.75em} &&&
	\\	\sync						& \lossysync					& \fifo	(Empty)						& \fifo (Full)
	\\	\begin{tikzpicture}[baseline]
	\node[reobnode]		(A) [label=below:$A$] {}; 
	\node[reobnode]		(B) [right of=A, label=below:$B$] {};
	\draw[sync]			(A) to node {} (B);
\end{tikzpicture}	& \begin{tikzpicture}[baseline]
	\node[reobnode]		(A) [label=below:$A$] {}; 
	\node[reobnode]		(B) [right of=A, label=below:$B$] {};
	\draw[lossysync]	(A) to node {} (B);
\end{tikzpicture}	& \ \begin{tikzpicture}[baseline]
	\node[reobnode]		(A) [label=below:$A$] {};
	\node[reobnode]		(B) [right of=A, label=below:$B$] {};
	\draw[emptyfifo]	(A) to node {} (B);
\end{tikzpicture}	& \begin{tikzpicture}[baseline]
	\node[reobnode]		(A) [label=below:$A$] {};
	\node[reobnode]		(B) [right of=A, label=below:$B$] {};	
	\draw[fullfifo]		(A) to node {} (B);
\end{tikzpicture}
	\\	\vspace{-.75em} &&&
	\\	\tablefill{.75em} $\tuple{\{ A , B \}, \{ \tuple{A , B} \}}$
	&	$\tuple{\{ A , B \}, \{ \tuple{A , B} \}}$
	&	\multicolumn{2}{c|}{$\tuple{\{ A , B \}, \{ \tuple{A , B} \}}$}
	\\	\hline
	\end{tabular}
	}
	\caption{Pictorial representation and formal definition of the structure of \sync, \lossysync, and \fifo.}
	\label{fig:connstruct:prim}
\end{figure}
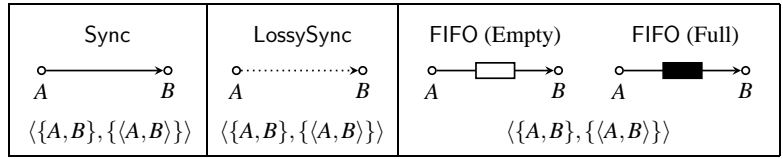

To illustrate the previous definition, Figure \ref{fig:connstruct:prim} shows pictorial representations and formal definitions of the structures of three common (binary) primitives. Because the pictorial representations of these primitives may give away some hints about their behavior, we discuss these informally here; the formal definitions appear later in this section. The \sync primitive consists of an input node and an output node. Data items flow through this primitive only if both of its nodes have pending I/O requests. The \lossysync primitive behaves similarly, but loses a data item if its input node has a pending write request while its output node has no pending take request. In contrast to the previous two primitives, connectors can have \emph{buffers} to store data items in. Such connectors exhibit different states, while the internal configuration of \sync and \lossysync always stays the same. For instance, the \fifo primitive consists of an input node, an output node, and a buffer of size 1. In its \textsc{empty} state, a write request on the input node of \fifo causes a data item to flow into its buffer|i.e., this buffer becomes full|while a take request on its output node remains pending. Conversely, in its \textsc{full} state, a write request on its input node remains pending, while a take request on its output node causes a data item to flow from the buffer to this output node|i.e., the buffer becomes empty. Finally, note the equality of the formal definitions of the structures of \sync, \lossysync, and \fifo: $\tuple{\{ A , B \}, \{ \tuple{A , B} \}}$. In general, all primitives that route data items from a single input node to a single output node have this structure (up to node renaming).

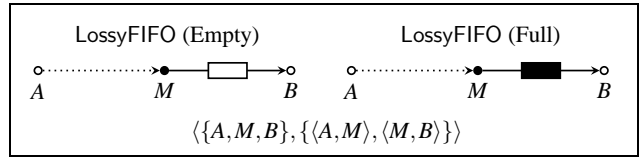
\begin{figure}[t]
	\centering
	\scalebox{.8}{
	\begin{tabular}{|cc|}
		\hline &
	\\	\vspace{-1.75em} &
	\\	\lossyfifo (Empty)					& \lossyfifo (Full)
	\\	\begin{tikzpicture}[baseline]
	\node[reobnode]		(A) [label=below:$A$] {}; 
	\node[reonode]		(M) [right of=A, label=below:$M$] {};
	\node[reobnode]		(B) [right of=M, label=below:$B$] {};
	\draw[lossysync]	(A) to node {} (M);
	\draw[emptyfifo]	(M) to node {} (B);
\end{tikzpicture}	& \begin{tikzpicture}[baseline]
	\node[reobnode]		(A) [label=below:$A$] {}; 
	\node[reonode]		(M) [right of=A, label=below:$M$] {};
	\node[reobnode]		(B) [right of=M, label=below:$B$] {};
	\draw[lossysync]	(A) to node {} (M);
	\draw[fullfifo]		(M) to node {} (B);
\end{tikzpicture}
	\\	\vspace{-.75em} &
	\\	\multicolumn{2}{|c|}{\tablefill{.75em} $\tuple{\{ A , M , B \}, \{\tuple{A , M} , \tuple{M , B} \}}$}
	\\	\hline
	\end{tabular}
	}
	\caption{Pictorial representation and formal definition of the structure of \lossyfifo.}
	\label{fig:connstruct:lossyfifo}
\end{figure}

We can construct complex connectors from simpler constituents (e.g., instances of primitives) using \emph{composition}. Connector structures $\connstruct_1$ and $\connstruct_2$ can compose if each of their shared nodes serves as an input node in $\connstruct_1$ and as an output node in $\connstruct_2$ or vice versa. To compose such compatible connector structures, we merge their sets of nodes and $\iomap$ relations.
\begin{definition}
	[Composition of connector structures]
	Let $\connstruct_1 = \tuple{\reonodeset_1 , \iomap_1}$ and $\connstruct_2 = \tuple{\reonodeset_2 , \iomap_2}$ be connector structures such that \emph{$\reonodeset_1 \cap \reonodeset_2 =  (\texttt{input}_{\connstruct_1} \cap \texttt{output}_{\connstruct_2}) \cup (\texttt{input}_{\connstruct_2} \cap \texttt{output}_{\connstruct_1})$}. Their composition, denoted $\connstruct_1 \connstructcomp \connstruct_2$, is a connector structure defined as:%
	\begin{center}
		$\connstruct_1 \connstructcomp \connstruct_2 = \tuple{\reonodeset_1 \cup \reonodeset_2 , \iomap_1 \cup \iomap_2}$
	\end{center}
\end{definition}
\noindent To illustrate the previous definition, Figure \ref{fig:connstruct:lossyfifo} shows the pictorial representation and formal definition of the structure of \lossyfifo, a connector composed of \lossysync and \fifo. The \lossyfifo connector consists of one input node, one internal node, and one output node. Similar to \fifo, the \lossyfifo connector exhibits the states \textsc{empty} and \textsc{full}. Informally, in the \textsc{empty} state, a write request on the input node of \lossyfifo \emph{always} causes a data item to flow into its buffer, while a take request on its output node remains pending. In the \textsc{full} state, a write request on its input node \emph{always} causes a data item to flow from its input node towards its buffer, but gets lost before reaching its internal node; a take request on its output node causes a data item to flow from its buffer to this output node.

Later in this paper, we use the possible compatibility between two connector structures as a well-formedness condition to define the composition of various \emph{comprehensive} models|models that combine a connector structure with a behavioral model|of Reo connectors.

\newcommand{\col}							{\kappa}
\newcommand{\coloring}						{c}
\newcommand{\coloringcomp}					{\cup}
\newcommand{\coloringtable}					{T}
\newcommand{\coloringtablecomp}				{\cdot}
\newcommand{\coloringtablemap}				{S}
\newcommand{\coloringtablemapcomp}			{\odot}
\newcommand{\coloringtableindex}			{\lambda}
\newcommand{\coloringtableindexset}			{\Lambda}
\newcommand{\coloringtableindexuniverse}	{\mbox{\upshape\textsf{Index}}}
\newcommand{\colsymflow}					{\, {\rule[0.5ex]{0.84cm}{1pt}} \,}
\newcommand{\colsymnoflow}					{\, {\hdashrule[0.5ex]{0.96cm}{1pt}{0.12cm}\hspace{-0.12cm}} \,}
\newcommand{\coluniverse}					{\mbox{\upshape\textsf{Color}}}

\newcommand{\nextfun}						{\eta}
\newcommand{\nextfuncomp}					{\otimes}

\newcommand{\nextfuninit}					{\epsilon}
\newcommand{\nextfuninitcomp}				{\nextfuncomp}

\newcommand{\conncolinit}		{\conncol}
\newcommand{\conncolinitcomp}	{\conncolcomp}

\subsection{Coloring Models}
\label{sect:prelim:cc}

\begin{figure}[t]
	\centering
	\scalebox{.8}{
	\begin{tabular}{|c|c|c|}
		\hline &&
	\\	\vspace{-1.75em} &&
	\\	\begin{tabular}[t]{@{}c@{}}
			\sync
		\\	\begin{tikzpicture}[baseline]
	\node[reobnode]		(A) [label=below:$A$] {}; 
	\node[reobnode]		(B) [right of=A, label=below:$B$] {};
	\draw[sync]			(A) to node {} (B);
\end{tikzpicture}
		\\	\begin{tikzpicture}[baseline]
	\node[invis]						(A1) [label=left:$c_1$] {};
	\node[invis, node distance=1.05cm]	(AB1) [right of=A1] {};
	\node[invis]						(B1) [right of=A1, label=right:\color{white}$c_1$] {};
	\draw[flow]							(A1) to node {} (B1);
	
	\node[invis, node distance=0.35cm]	(A2) [below of=A1, label=left:$c_4$] {};
	\node[invis, node distance=1.05cm]	(AB2) [right of=A2] {};
	\node[invis]						(B2) [right of=A2, label=right:\color{white}$c_4$] {};
	\draw[noflow]						(A2) to node {} (B2);
\end{tikzpicture}
		\end{tabular}
	&	\begin{tabular}[t]{@{}c@{}}
			\lossysync
		\\	\begin{tikzpicture}[baseline]
	\node[reobnode]		(A) [label=below:$A$] {}; 
	\node[reobnode]		(B) [right of=A, label=below:$B$] {};
	\draw[lossysync]	(A) to node {} (B);
\end{tikzpicture}
		\\	\begin{tikzpicture}[baseline]
	\node[invis]						(A1) [label=left:$c_1$] {};
	\node[invis, node distance=1.05cm]	(AB1) [right of=A1] {};
	\node[invis]						(B1) [right of=A1, label=right:\color{white}$c_1$] {};
	\draw[flow]							(A1) to node {} (B1);
	
	\node[invis, node distance=0.35cm]	(A2) [below of=A1, label=left:$c_2$] {};
	\node[invis, node distance=1.05cm]	(AB2) [right of=A2] {};
	\node[invis]						(B2) [right of=A2, label=right:\color{white}$c_2$] {};
	\draw[flow]							(A2) to node {} (AB2);
	\draw[noflow]						(AB2) to node {} (B2);
	
	\node[invis, node distance=0.35cm]	(A3) [below of=A2, label=left:$c_4$] {};
	\node[invis, node distance=1.05cm]	(AB3) [right of=A3] {};
	\node[invis]						(B3) [right of=A3, label=right:\color{white}$c_4$] {};
	\draw[noflow]						(A3) to node {} (B3);
\end{tikzpicture}
		\end{tabular}
	&	\begin{tabular}[t]{@{}c@{\hspace{-1em}}c@{\hspace{-1em}}}
			\fifo (Empty)					& \fifo (Full)
		\\	\begin{tikzpicture}[baseline]
	\node[reobnode]		(A) [label=below:$A$] {};
	\node[reobnode]		(B) [right of=A, label=below:$B$] {};
	\draw[emptyfifo]	(A) to node {} (B);
\end{tikzpicture}	& \begin{tikzpicture}[baseline]
	\node[reobnode]		(A) [label=below:$A$] {};
	\node[reobnode]		(B) [right of=A, label=below:$B$] {};	
	\draw[fullfifo]		(A) to node {} (B);
\end{tikzpicture}
		\\	\begin{tikzpicture}[baseline]
	\node[invis]						(A1) [label=left:$c_2$] {};
	\node[invis, node distance=1.05cm]	(AB1) [right of=A1] {};
	\node[invis]						(B1) [right of=A1, label=right:\color{white}$c_2$] {};
	\draw[flow]							(A1) to node {} (AB1);
	\draw[noflow]						(AB1) to node {} (B1);
	
	\node[invis, node distance=0.35cm]	(A2) [below of=A1, label=left:$c_4$] {};
	\node[invis, node distance=1.05cm]	(AB2) [right of=A2] {};
	\node[invis]						(B2) [right of=A2, label=right:\color{white}$c_4$] {};
	\draw[noflow]						(A2) to node {} (B2);
\end{tikzpicture}	& \begin{tikzpicture}[baseline]
	\node[invis]						(A1) [label=left:$c_3$] {};
	\node[invis, node distance=1.05cm]	(AB1) [right of=A1] {};
	\node[invis]						(B1) [right of=A1, label=right:\color{white}$c_3$] {};
	\draw[noflow]						(A1) to node {} (AB1);
	\draw[flow]							(AB1) to node {} (B1);
	
	\node[invis, node distance=0.35cm]	(A2) [below of=A1, label=left:$c_4$] {};
	\node[invis, node distance=1.05cm]	(AB2) [right of=A2] {};
	\node[invis]						(B2) [right of=A2, label=right:\color{white}$c_4$] {};
	\draw[noflow]						(A2) to node {} (B2);
\end{tikzpicture}
		\end{tabular}
	\\	\vspace{-.75em} &&
	\\	\tablefill{9em} $\begin{array}[t]{@{}l@{\:}c@{\:}l@{}}
			\coloringtablemap & = & \left\{ \begin{array}{@{}c@{}}
				\syncindex \mapsto \{ \coloring_1 , \coloring_4 \}
			\end{array} \right\}
		\\	\vspace{1.5em}
		\\	\nextfun & = & \left\{ \begin{array}{@{}l@{}}
				\tuple{\syncindex , \coloring_1} \mapsto \syncindex \; ,
			\\	\tuple{\syncindex , \coloring_4} \mapsto \syncindex
			\end{array} \right\}
		\\	\vspace{.5em}
		\\	\coloringtableindex_0 & = & \syncindex
		\end{array}$
	&	$\begin{array}[t]{@{}l@{\:}c@{\:}l@{}}
			\coloringtablemap & = & \left\{ \begin{array}{@{}c@{}}
				\lossysyncindex \mapsto \{ \coloring_1 , \coloring_2 , \coloring_4 \}
			\end{array} \right\}
		\\	\vspace{.75em}
		\\	\nextfun & = & \left\{ \begin{array}{@{}l@{}}
				\tuple{\lossysyncindex , \coloring_1} \mapsto \lossysyncindex \; ,
			\\	\tuple{\lossysyncindex , \coloring_2} \mapsto \lossysyncindex \; ,
			\\	\tuple{\lossysyncindex , \coloring_4} \mapsto \lossysyncindex \;
			\end{array} \right\}
		\\	\vspace{0em}
		\\	\coloringtableindex_0 & = & \lossysyncindex
		\end{array}$
	&	$\begin{array}[t]{@{}l@{\:}c@{\:}l@{}}
			\coloringtablemap & = & \left\{ \begin{array}{@{}l@{}}
				\fifoemptyindex \mapsto \{ \coloring_2 , \coloring_4 \} \; ,
			\\	\fifofullindex \mapsto \{ \coloring_3 , \coloring_4 \}
			\end{array} \right\}
		\\	\vspace{-.75em}
		\\	\nextfun & = & \left\{ \begin{array}{@{}l@{}}
				\tuple{\fifoemptyindex , \coloring_2} \mapsto \fifofullindex \; ,
			\\	\tuple{\fifoemptyindex , \coloring_4} \mapsto \fifoemptyindex \; ,
			\\	\vspace{-.75em}
			\\	\tuple{\fifofullindex , \coloring_3} \mapsto \fifoemptyindex \; ,
			\\	\tuple{\fifofullindex , \coloring_4} \mapsto \fifofullindex
			\end{array} \right\}
		\\	\vspace{-.75em}
		\\	\coloringtableindex_0 & = & \fifoemptyindex
		\end{array}$
	\\	\hline
	\end{tabular}
	}
	\caption{Colorings, CTMs, and initialized next functions of \sync, \lossysync, and \fifo.}
	\label{fig:nextfun:prim}
\end{figure}
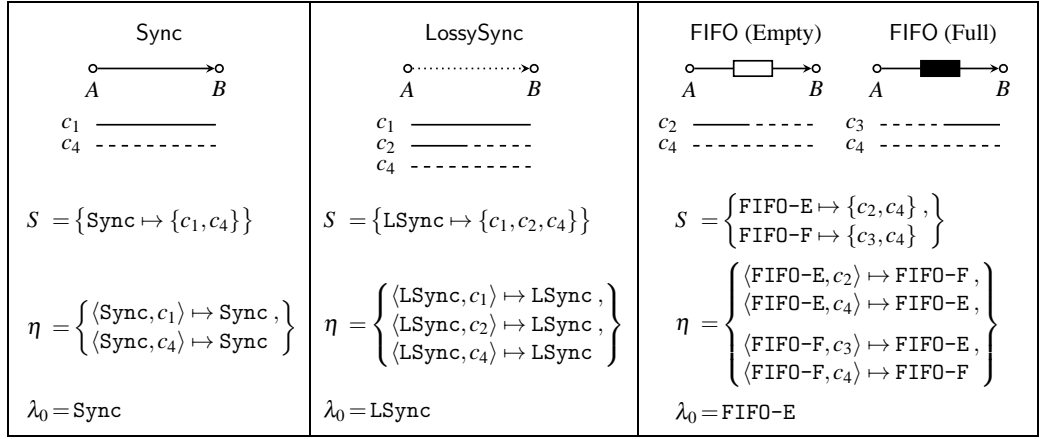

We proceed with \emph{coloring models} \cite{clarke07,costa10} as the first semantic model we discuss. Coloring models work by marking nodes of a connector with \emph{colors} that specify whether data items flow through these nodes or not. Depending on the number of colors, different models with different levels of expressiveness arise. In this paper, we assume a total of two colors: the \emph{flow color} $\colsymflow$ (data items can flow through the nodes it marks) and the \emph{no-flow color} $\colsymnoflow$ (data items cannot flow through the nodes it marks).%
\footnote{%
	Two colors yield the \emph{2-coloring model}. Alternatively, the \emph{3-coloring model} features three different colors to mark nodes with. Although many believe that the 3-coloring model has a higher degree of expressiveness than the 2-coloring model, a recent investigation \cite{jongmans11} suggests otherwise.
} To describe a single behavior alternative of a connector in a given state, we define \emph{colorings}: maps from sets of nodes to sets of colors, which assign to each node in the set a color that indicates whether this node fires (or not) in the behavior that the coloring describes. We collect all behavior alternatives of a connector in sets of colorings called \emph{coloring tables}.
\begin{definition}
	[Colors \cite{clarke07}]
	\label{def:coluniverse}
	$\coluniverse = \setleft \colsymflow , \colsymnoflow \setright$ is a set of colors.
\end{definition}
\begin{definition}
	[Coloring \cite{clarke07}]
	\label{def:coloring}
	Let $\reonodeset \subseteq \reonodeuniverse$. A coloring $\coloring$ over $\reonodeset$ is a map $\reonodeset \rightarrow \coluniverse$.
\end{definition}
\begin{definition}
	[Coloring table \cite{clarke07}]
	\label{def:coloringtable}
	Let $\reonodeset \subseteq \reonodeuniverse$. A coloring table $\coloringtable$ over $\reonodeset$ is a set $\setleft \reonodeset \rightarrow \coluniverse \setright$ of colorings over $\reonodeset$.
\end{definition}
\noindent To accommodate connectors that exhibit different behavior in different states (e.g., connectors with buffers to store data items in), we use \emph{coloring table maps} (CTM): maps from sets of \emph{indexes} (representing the states of a connector) to sets of coloring tables (describing the allowed behavior alternatives in these states).%
\footnote{%
	In \cite{costa10}, Costa calls coloring table maps \emph{indexed sets of coloring tables}.
} Subsequently, to model the change of state a connector incurs when (some of) its nodes fire, we use \emph{next functions}. The next function of a connector maps an index $\coloringtableindex$ in the domain $\coloringtableindexset$ of a CTM $\coloringtablemap$ and a coloring in the coloring table to which $\coloringtablemap$ maps $\coloringtableindex$ to some index in $\coloringtableindexset$ (possibly the same $\coloringtableindex$).
\begin{definition}
	[Universe of indexes]
	\label{def:coloringtableindexuniverse}
	$\coloringtableindexuniverse$ is the set of indexes.
\end{definition}
\begin{definition}
	[Coloring table map \cite{costa10}]
	\label{def:coloringtablemap}
	Let $\reonodeset \subseteq \reonodeuniverse$ and $\coloringtableindexset \subseteq \coloringtableindexuniverse$. A coloring table map $\coloringtablemap$ over $[\reonodeset , \coloringtableindexset]$ is a total map $\coloringtablemap : \coloringtableindexset \rightarrow 2^{\{ \reonodeset \rightarrow \mbox{\scriptsize $\coluniverse$} \}}$ from indexes to coloring tables over $\reonodeset$.
\end{definition}
\begin{definition}
	[Next function \cite{costa10}]
	\label{def:nextfun}
	Let $\coloringtablemap$ be a CTM over $[\reonodeset , \coloringtableindexset]$. A next function $\nextfun$ over $\coloringtablemap$ is a partial map $\coloringtableindexset \times \setleft \reonodeset \rightarrow \coluniverse \setright \rightharpoonup \coloringtableindexset$ from [index, coloring]-pairs to indexes such that $[\coloringtableindex , \coloring \mapsto \coloringtableindex'] \in \nextfun$ iff $\coloring \in \coloringtablemap(\coloringtableindex)$.
\end{definition}
\noindent Next, we slightly extend the connector coloring framework as presented in \cite{clarke07,costa10} by introducing \emph{initialized next functions}. Suppose a CTM $\coloringtablemap$ over $[\reonodeset , \coloringtableindexset]$ and a next function $\nextfun$ over $\coloringtablemap$: an \emph{initialization} of $\nextfun$|formally a pair of a next function and an index|associates $\nextfun$ with some $\coloringtableindex_0 \in \coloringtableindexset$. This $\coloringtableindex_0$ represents the initial state of the connector whose behavior $\nextfun$ models.
\begin{definition}
	[Initialized next function]
	\label{def:nextfuninit}
	Let $\coloringtablemap$ be a CTM over $[\reonodeset , \coloringtableindexset]$. An initialized next function $\nextfuninit$ over $\coloringtablemap$ is a pair $\tuple{\nextfun , \coloringtableindex_0}$ with $\nextfun$ a next function over $\coloringtablemap$ and $\coloringtableindex_0 \in \coloringtableindexset$.
\end{definition}
\noindent Finally, we join coloring models|i.e., initialized next functions, which comprehensively describe the behavior of circuits|and connector structures in \emph{$\nextfuninit$-connectors}: complete formal models of connectors.%
\begin{definition}
	[$\nextfuninit$-connector]
	\label{def:conncolinit}
	Let $\reonodeset \subseteq \reonodeuniverse$ and let $\coloringtablemap$ be a CTM over $[\reonodeset , \coloringtableindexset]$. An $\nextfuninit$-connector $\conncolinit$ over $[\reonodeset , \coloringtablemap]$ is a pair $\tuple{\connstruct , \nextfuninit}$ with $\connstruct = \tuple{\reonodeset , \iomap}$ a connector structure and $\nextfuninit$ an initialized next function over $\coloringtablemap$. 
\end{definition}
\noindent To illustrate the previous definitions, Figure \ref{fig:nextfun:prim} shows the coloring models that describe the behavior of \sync, \lossysync, and \fifo, whose structures we depicted in Figure \ref{fig:connstruct:prim}.%
\footnote{%
	\label{footnote:index}%
	Because a composed connector may consist of multiple instances of the same (primitive) connector, we should actually use indexes that enable distinguishing between such instances. For example, rather than just ``\syncindex'', we could extend this index with a distinctive subscript, such as the set of nodes that the respective \sync primitive connects (e.g., ``$\syncindex_{A,B}$'') or, alternatively, an integer (e.g., ``$\syncindex_1$''). Throughout this paper, however, we abstract from such subscripts for notational convenience.
}

When we compose two connectors that have coloring models as formal semantics, we can compute the coloring model of the composed connector by composing the coloring models of its constituents. We describe this composition process in a bottom-up fashion. First, to compose two \emph{compatible colorings}|i.e., colorings that assign the same colors to their shared nodes|we merge the domains of these colorings and map each node $\reonode$ in the resulting set to the color that one of the colorings assigns to $\reonode$. The composition of two coloring tables then comprises the computation of a new coloring table that contains the pairwise compositions of the compatible colorings in the two individual coloring tables.%
\begin{definition}
	[Composition of colorings \cite{clarke07}]
	\label{def:coloringcomp}
	Let $\coloring_1$ and $\coloring_2$ be colorings over $\reonodeset_1$ and $\reonodeset_2$ such that $\coloring_1(\reonode) = \coloring_2(\reonode)$ for all $\reonode \in \reonodeset_1 \cap \reonodeset_2$. Their composition, denoted by $\coloring_1 \coloringcomp \coloring_2$, is a coloring over $\reonodeset_1 \cup \reonodeset_2$ defined as:
	\begin{center}$
		\coloring_1 \coloringcomp \coloring_2 = \left \{ \reonode \mapsto \col \left| \begin{array}{@{}l@{}}
				\reonode \in \reonodeset_1 \cup \reonodeset_2 \mbox{ and } \col = \left( \begin{array}{@{}l@{\enspace}l@{}}
					\coloring_1(\reonode) & \mbox{if } \reonode \in \reonodeset_1
				\\	\coloring_2(\reonode) & \mbox{otherwise}
				\end{array} \right)
			\end{array} \right. \right\}
	$\end{center}
\end{definition}
\begin{definition}
	[Composition of coloring tables \cite{clarke07}]
	\label{def:coloringtablecomp}
	Let $\coloringtable_1$ and $\coloringtable_2$ be coloring tables over $\reonodeset_1$ and $\reonodeset_2$. Their composition, denoted by $\coloringtable_1 \coloringtablecomp \coloringtable_2$, is a coloring table over $\reonodeset_1 \cup \reonodeset_2$ defined as:
	\begin{center}$
		\coloringtable_1 \coloringtablecomp \coloringtable_2 = \setleft \coloring_1 \coloringcomp \coloring_2 \setbar \coloring_1 \in \coloringtable_1 \mbox{ and } \coloring_2 \in \coloringtable_2 \mbox{ and } \coloring_1(\reonode) = \coloring_2(\reonode) \mbox{ for all } \reonode \in \reonodeset_1 \cap \reonodeset_2 \setright
	$\end{center}
\end{definition}
\noindent Next, the composition of two CTMs comprises the computation of a new CTM that maps each pair of indexes in the Cartesian product of the domains of the two individual CTMs to the composition of the coloring tables to which these CTMs map the indexes in the pair. We define the composition of two next functions in terms of the Cartesian product, the composition of colorings, and the composition of CTMs.%
\begin{definition}
	[Composition of CTMs \cite{costa10}]
	\label{def:coloringtablemapcomp}
	Let $\coloringtablemap_1$ and $\coloringtablemap_2$ be CTMs over $[\reonodeset_1 , \coloringtableindexset_1]$ and $[\reonodeset_2 , \coloringtableindexset_2]$. Their composition, denoted by $\coloringtablemap_1 \coloringtablemapcomp \coloringtablemap_2$, is a CTM over $[\reonodeset_1 \cup \reonodeset_2 , \coloringtableindexset_1 \times \coloringtableindexset_2]$ defined as:
	\begin{center}$
		\coloringtablemap_1 \coloringtablemapcomp \coloringtablemap_2 = \setleft \tuple{\coloringtableindex_1 , \coloringtableindex_2} \mapsto \coloringtablemap_1(\coloringtableindex_1) \coloringtablecomp \coloringtablemap_2(\coloringtableindex_2) \setbar \coloringtableindex_1 \in \coloringtableindexset_1 \mbox{ and } \coloringtableindex_2 \in \coloringtableindexset_2 \setright
	$\end{center}
\end{definition}
\begin{definition}
	[Composition of next functions \cite{costa10}]
	\label{def:nextfuncomp}
	Let $\nextfun_1$ and $\nextfun_2$ be next functions over $\coloringtablemap_1$ and $\coloringtablemap_2$ with $\coloringtablemap_1$ and $\coloringtablemap_2$ defined over $[\reonodeset_1 , \coloringtableindexset_1]$ and $[\reonodeset_2 , \coloringtableindexset_2]$. Their composition, denoted by $\nextfun_1 \nextfuncomp \nextfun_2$, is a next function over $\coloringtablemap_1 \coloringtablemapcomp \coloringtablemap_2$ defined as:
	\begin{center}$
		\begin{array}{@{}l@{\;}c@{\;}l@{}}
			\nextfun_1 \nextfuncomp \nextfun_2 & = & \left\{ \left. \begin{array}{@{}c@{\;}}
				\tuple{\coloringtableindex_1 , \coloringtableindex_2} , \coloring_1 \coloringcomp \coloring_2 
			\\	\rotatebox[origin=c]{270}{$\mapsto$}
			\\	\tuple{\nextfun_1(\coloringtableindex_1 , \coloring_1) , \nextfun_2(\coloringtableindex_2 , \coloring_2)}
			\end{array} \right| \begin{array}{@{\;}c@{}}
				\tuple{\coloringtableindex_1 , \coloringtableindex_2} \in \coloringtableindexset_1 \times \coloringtableindexset_2
			\\	\mbox{and}
			\\	\coloring_1 \coloringcomp \coloring_2 \in (\coloringtablemap_1 \coloringtablemapcomp \coloringtablemap_2)(\tuple{\coloringtableindex_1 , \coloringtableindex_2})
			\end{array} \right\}
		\end{array}
	$\end{center}
\end{definition}
\noindent We define the composition of initialized next functions in terms of the composition of next functions and take the pair of the initial states as the initial state of the composition.
\begin{definition}
	[Composition of initialized next functions]
	\label{def:nextfuninitcomp}
	Let $\nextfuninit_1 = \tuple{\nextfun_1 , \coloringtableindex_0^1}$ and $\nextfuninit_2 = \tuple{\nextfun_2 , \coloringtableindex_0^2}$ be initialized next functions over $\coloringtablemap_1$ and $\coloringtablemap_2$. Their composition, denoted by $\nextfuninit_1 \nextfuninitcomp \nextfuninit_2$, is an initialized next function over $\coloringtablemap_1 \coloringtablemapcomp \coloringtablemap_2$ defined as:
	\begin{center}$
		\nextfuninit_1 \nextfuninitcomp \nextfuninit_2 = \tuple{\nextfun_1 \nextfuncomp \nextfun_2 , \tuple{\coloringtableindex_0^1 , \coloringtableindex_0^2}}
	$\end{center}
\end{definition}
\noindent Finally, we define the composition of $\nextfuninit$-connectors in terms of the composition of connector structures and the composition of initialized next functions.
\begin{definition}
	[Composition of $\nextfuninit$-connectors]
	\label{def:conncolinitcomp}
	Let $\conncolinit_1 = \tuple{\connstruct_1 , \nextfuninit_1}$ and $\conncolinit_2 = \tuple{\connstruct_2 , \nextfuninit_2}$ be $\nextfuninit$-connectors over $[\reonodeset_1 , \coloringtablemap_1]$ and $[\reonodeset_2 , \coloringtablemap_2]$ such that $\connstruct_1 \connstructcomp \connstruct_2$ is defined. Their composition, denoted by $\conncolinit_1 \conncolinitcomp \conncolinit_2$, is an $\nextfuninit$-connector over $[\reonodeset_1 \cup \reonodeset_2 , \coloringtablemap_1 \coloringtablemapcomp \coloringtablemap_2]$ defined as:
	\begin{center}$
		\conncolinit_1 \conncolinitcomp \conncolinit_2 = \tuple{\connstruct_1 \connstructcomp \connstruct_2 , \nextfuninit_1 \nextfuninitcomp \nextfuninit_2}
	$\end{center}
\end{definition}
\noindent To illustrate the previous definitions, Figure \ref{fig:nextfun:lossyfifo} shows a coloring model that describes the behavior of \lossyfifo, whose structure we depicted in Figure \ref{fig:connstruct:lossyfifo}. In this figure, the index of a coloring specifies its origin: a coloring $\coloring_{ij}$ results from composing $\coloring_i$ (of \lossysync) with $\coloring_j$ (of \fifo) in Figure \ref{fig:nextfun:prim}.%
\footnote{%
	Note that this composed coloring model of \lossyfifo does \emph{not} describe its intended behavior as outlined near the end of the previous subsection: coloring $\coloring_{24}$, allowed in the \textsc{empty} state according to the composed model, describes the loss of a data item between $A$ and $M$. We, however, consider this inadmissible behavior in the \textsc{empty} state. Clarke et al. recognize this inconsistency between intuition and formal practice in \cite{clarke07} and call it the problem of describing \emph{context-sensitive} connectors. We refer the reader interested in the challenges that context-sensitive connectors entail to \cite{bonsangue09,clarke07,costa10,jongmans11}.
}

\begin{figure}[t]
	\centering
	\scalebox{.8}{
	\begin{tabular}{|c|}
		\hline
	\\	\vspace{-1.75em}
	\\	\begin{tabular}[t]{@{}c@{}c@{}}
			\lossyfifo (Empty)& \lossyfifo (Full)
		\\	\begin{tikzpicture}[baseline]
	\node[reobnode]		(A) [label=below:$A$] {}; 
	\node[reonode]		(M) [right of=A, label=below:$M$] {};
	\node[reobnode]		(B) [right of=M, label=below:$B$] {};
	\draw[lossysync]	(A) to node {} (M);
	\draw[emptyfifo]	(M) to node {} (B);
\end{tikzpicture}	& \begin{tikzpicture}[baseline]
	\node[reobnode]		(A) [label=below:$A$] {}; 
	\node[reonode]		(M) [right of=A, label=below:$M$] {};
	\node[reobnode]		(B) [right of=M, label=below:$B$] {};
	\draw[lossysync]	(A) to node {} (M);
	\draw[fullfifo]		(M) to node {} (B);
\end{tikzpicture}
		\\	\begin{tikzpicture}[baseline]
	\node[invis]						(A1) [label=left:$c_{12}$] {};
	\node[invis, node distance=1.05cm]	(AM1) [right of=A1] {};
	\node[invis]						(M1) [right of=A1] {};
	\node[invis, node distance=1.05cm]	(MB1) [right of=M1] {};
	\node[invis]						(B1) [right of=M1, label=right:\color{white}$c_{12}$] {};
	\draw[flow]							(A1) to node {} (MB1);
	\draw[noflow]						(MB1) to node {} (B1);
	
	\node[invis, node distance=0.35cm]	(A2) [below of=A1, label=left:$c_{24}$] {};
	\node[invis, node distance=1.05cm]	(AM2) [right of=A2] {};
	\node[invis]						(M2) [right of=A2] {};
	\node[invis, node distance=1.05cm]	(MB2) [right of=M2] {};
	\node[invis]						(B2) [right of=M2, label=right:\color{white}$c_{24}$] {};
	\draw[flow]							(A2) to node {} (AM2);
	\draw[noflow]						(AM2) to node {} (B2);

	\node[invis, node distance=0.35cm]	(A3) [below of=A2, label=left:$c_{44}$] {};
	\node[invis, node distance=1.05cm]	(AM3) [right of=A3] {};
	\node[invis]						(M3) [right of=A3] {};
	\node[invis, node distance=1.05cm]	(MB3) [right of=M3] {};
	\node[invis]						(B3) [right of=M3, label=right:\color{white}$c_{44}$] {};
	\draw[noflow]						(A3) to node {} (B3);
\end{tikzpicture}	& \begin{tikzpicture}[baseline]
	\node[invis]						(A1) [label=left:$c_{23}$] {};
	\node[invis, node distance=1.05cm]	(AM1) [right of=A1] {};
	\node[invis]						(M1) [right of=A1] {};
	\node[invis, node distance=1.05cm]	(MB1) [right of=M1] {};
	\node[invis]						(B1) [right of=M1, label=right:\color{white}$c_{23}$] {};
	\draw[flow]							(A1) to node {} (AM1);
	\draw[noflow]						(AM1) to node {} (MB1);
	\draw[flow]							(MB1) to node {} (B1);
	
	\node[invis, node distance=0.35cm]	(A2) [below of=A1, label=left:$c_{24}$] {};
	\node[invis, node distance=1.05cm]	(AM2) [right of=A2] {};
	\node[invis]						(M2) [right of=A2] {};
	\node[invis, node distance=1.05cm]	(MB2) [right of=M2] {};
	\node[invis]						(B2) [right of=M2, label=right:\color{white}$c_{24}$] {};
	\draw[flow]							(A2) to node {} (AM2);
	\draw[noflow]						(AM2) to node {} (B2);

	\node[invis, node distance=0.35cm]	(A3) [below of=A2, label=left:$c_{43}$] {};
	\node[invis, node distance=1.05cm]	(AM3) [right of=A3] {};
	\node[invis]						(M3) [right of=A3] {};
	\node[invis, node distance=1.05cm]	(MB3) [right of=M3] {};
	\node[invis]						(B3) [right of=M3, label=right:\color{white}$c_{43}$] {};
	\draw[noflow]						(A3)	to node {} (MB3);
	\draw[flow]							(MB3) to node {} (B3);
	
	\node[invis, node distance=0.35cm]	(A4) [below of=A3, label=left:$c_{44}$] {};
	\node[invis, node distance=1.05cm]	(AM4) [right of=A4] {};
	\node[invis]						(M4) [right of=A4] {};
	\node[invis, node distance=1.05cm]	(MB4) [right of=M4] {};
	\node[invis]						(B4) [right of=M4, label=right:\color{white}$c_{44}$] {};
	\draw[noflow]						(A4) to node {} (B4);
\end{tikzpicture}
		\end{tabular}
	\\	\vspace{-.75em}
	\\	\tablefill{8em} $\begin{array}[t]{@{}l@{\:}c@{\:}l@{}}
			\coloringtablemap & = & \left\{ \begin{array}{@{}l@{}}
				\lossyfifoemptyindex \mapsto \{ \coloring_{12} , \coloring_{24} , \coloring_{44} \} \; , \; \lossyfifofullindex \mapsto \{ \coloring_{23} , \coloring_{24} , \coloring_{43} , \coloring_{44} \}
			\end{array} \right\}
		\\	\vspace{-.75em}
		\\	\nextfun & = & \left\{ \begin{array}{@{}l@{\;}c@{\;}l@{}}
				\tuple{\lossyfifoemptyindex , \coloring_{12}} \mapsto \lossyfifofullindex & , & \tuple{\lossyfifofullindex , \coloring_{23}} \mapsto \lossyfifoemptyindex \; ,
			\\	\tuple{\lossyfifoemptyindex , \coloring_{24}} \mapsto \lossyfifoemptyindex & , & \tuple{\lossyfifofullindex , \coloring_{24}} \mapsto \lossyfifofullindex ,
			\\	\tuple{\lossyfifoemptyindex , \coloring_{44}} \mapsto \lossyfifoemptyindex & , & \tuple{\lossyfifofullindex , \coloring_{43}} \mapsto \lossyfifoemptyindex \; ,
			\\	& & \tuple{\lossyfifofullindex , \coloring_{44}} \mapsto \lossyfifofullindex
			\end{array} \right\}
		\\	\vspace{-.75em}
		\\	\coloringtableindex_0 & = & \lossyfifoemptyindex
		\end{array}$
	\\	\hline
	\end{tabular}
	}
	\caption{Colorings, CTM, and next function of \lossyfifo. We abbreviate $\tuple{\lossysyncindex , \fifoemptyindex}$ by $\lossyfifoemptyindex$ and $\tuple{\lossysyncindex , \fifofullindex}$ by $\lossyfifofullindex$.}
	\label{fig:nextfun:lossyfifo}
\end{figure}

\newcommand{\ca}						{\alpha}
\newcommand{\caconstraint}				{g}
\newcommand{\caconstraintset}			{G}
\newcommand{\caconstraintuniverse}		{\mbox{\upshape\textsf{Constraint}}}
\newcommand{\cacomp}					{\bowtie}
\newcommand{\castate}					{q}
\newcommand{\castateset}				{Q}
\newcommand{\catransrel}				{R}

\newcommand{\dataitem}					{d}
\newcommand{\dataitemuniverse}			{\mbox{\upshape\textsf{Data}}}

\subsection{Constraint Automata}
\label{sect:prelim:ca}

We end this section with \emph{constraint automata} (CA) \cite{baier06} as the second semantic model we discuss. A CA consists of a (possibly singleton) set of states, which correspond one-to-one to the states of the connector whose behavior it models and a set of transitions between them; in contrast to standard automata, CA do not have accepting states. A transition of a CA carries a label that consists of two elements: a set of nodes and a \emph{data constraint}. The former, called a \emph{firing set}, describes which nodes fire simultaneously in the state the transition leaves from; the latter specifies the conditions that the content of the data items that flow through these firing nodes must satisfy. To define (the universe of) data constraints, we assume a universe of data items. This set contains every data item that we may send through a Reo connector.
\begin{definition}
	[Universe of data items]
	\label{def:dataitemuniverse}
	$\dataitemuniverse$ is the set of data items.
\end{definition}
\begin{definition}
	[Universe of data constraints \cite{baier06}]
	\label{def:caconstraintuniverse}
	$\caconstraintuniverse$ is the set of data constraints such that each $\caconstraint \in \caconstraintuniverse$ complies with the following grammar:
	\begin{center}
		$\caconstraint ::= \caconstraint \wedge \caconstraint \setbar \neg \caconstraint \setbar \top \setbar \#\reonode = \dataitem \mbox{ with } \reonode \in \reonodeuniverse \mbox{ and } \dataitem \in \dataitemuniverse$
	\end{center}	
\end{definition}
\noindent Informally, $\#\reonode$ means ``the data item that flows through $\reonode$'', while $\wedge$, $\neg$, and $\top$ have their usual meaning. For convenience, we also allow their derived Boolean operators such as $\vee$, $\Rightarrow$ (implication), etc., as syntactic sugar; we adopt $\#\reonode_1 = \#\reonode_2$ (with $\reonode_1 , \reonode_2 \in \reonodeuniverse$) as an abbreviation of $\bigvee_{\dataitem \in \mbox{\scriptsize $\dataitemuniverse$}} (\#\reonode_1 = \dataitem \wedge \#\reonode_2 = \dataitem)$. This gives us sufficient machinery to define CA. Recall that CA serve as operational models of connector behavior|their states correspond one-to-one to the states of a connector, while their transitions specify for each state when and what data items can flow through which nodes.

\begin{definition}
	[Constraint automaton \cite{baier06}]
	\label{def:ca}
	Let $\reonodeset \subseteq \reonodeuniverse$ and $\caconstraintset \subseteq \caconstraintuniverse$. A constraint automaton $\ca$ over $[\reonodeset , \caconstraintset]$ is a tuple $\tuple{\castateset , \catransrel , \castate_0}$ with $\castateset$ a set of states, $\catransrel \subseteq \castateset \times 2^{\reonodeset} \times \caconstraintset \times \castateset$ a transition relation, and $\castate_0 \in \castateset$ an initial state.
\end{definition}
\noindent As initialized next functions, CA comprehensively model circuit behavior. Similar to $\nextfuninit$-connectors, therefore, we introduce \emph{$\ca$-connectors}: pairs that consist of a connector structure and a CA.%
\begin{definition}
	[$\alpha$-connector]
	\label{def:connca}
	Let $\reonodeset \subseteq \reonodeuniverse$ and $\caconstraintset \subseteq \caconstraintuniverse$. An $\ca$-connector $\connca$ over $[\reonodeset , \caconstraintset]$ is a pair $\tuple{\connstruct , \ca}$ with $\connstruct = \tuple{\reonodeset , \iomap}$ a connector structure and $\ca = \tuple{\castateset , \catransrel , \castate_0}$ a CA over $[\reonodeset , \caconstraintset]$.
\end{definition}

\begin{figure}[t]
	\centering
	\scalebox{.8}{
	\begin{tabular}{|l|l|l|}
		\hline &&
	\\	\vspace{-2.25em} &&
	\\	\begin{tikzpicture}[baseline]
	\node[state] (Q) [pin={[pin edge={black, thick, dotted}]below:\syncindex}] {};
	\path[->] (Q) edge [in=15,		out=-15, thick, loop] node [right]	{\parbox{30.99086pt}{\footnotesize $\begin{array}{@{}c@{}}\{ A , B \} , \\ \#A = \#B\end{array}$}} (Q);
	\path[->] (Q) edge [in=-165,	out=165, thick, loop] node [left]	{\footnotesize $\emptyset , \top$}(Q);
\end{tikzpicture} & \begin{tikzpicture}[baseline]
	\node[state] (Q) [pin={[pin edge={black, thick, dotted}]below:\lossysyncindex}] {};
	\path[->] (Q) edge [in=15,		out=-15, thick, loop] node [right]	{\parbox{30.99086pt}{\footnotesize $\begin{array}{@{}c@{}}\{ A , B \} , \\ \#A = \#B\end{array}$}} (Q);
	\path[->] (Q) edge [in=105,		out=75, thick, loop] node [above]	{\footnotesize $\{ A \} , \top$} (Q);
	\path[->] (Q) edge [in=-165,	out=165, thick, loop] node [left]	{\footnotesize $\emptyset , \top$} (Q);
\end{tikzpicture} & \begin{tikzpicture}[baseline]
	\tikz[pin distance=.5cm];
	\node[state] 					(Q) [pin={[pin edge={black, thick, dotted}]below:\fifoemptyindex}] {};
	\node[state, node distance=3cm]	(R) [right of=Q, pin={[pin edge={black, thick, dotted}]below:\fifofullindex}] {};
	\path[->] (Q) edge [in=-165,	out=-15, thick]			node [below]	{\footnotesize $\{ A \} , \#A = \mbox{\upshape ``foo''}$} (R);
	\path[->] (Q) edge [in=-165,	out=165, thick, loop]	node [left]		{\footnotesize $\emptyset , \top$} (Q);
	\path[->] (R) edge [in=15,		out=165, thick]			node [above]	{\footnotesize $\{ B \} , \#B = \mbox{\upshape ``foo''}$} (Q);
	\path[->] (R) edge [in=15,		out=-15, thick, loop]	node [right]	{\footnotesize $\emptyset , \top$} (R);
\end{tikzpicture}
	\\	\hline
	\end{tabular}
	}
	\caption{Constraint automata of \sync, \lossysync, and \fifo with $\dataitemuniverse = \{ \mbox{\upshape ``foo''} \}$.}
	\label{fig:pa:prim}
\end{figure}

\noindent To illustrate the previous definitions, the CA of \sync, \lossysync, and \fifo appear in Figure \ref{fig:pa:prim}. For simplicity, we assume the universe of data items a singleton with the string ``foo'' as its sole element. In general, the CA of \fifo contains a distinct state for each data item in $\dataitemuniverse$ that may occupy its buffer. Note that rather than naming states symbolically (e.g., $q , p , q_0 , q_1, \ldots$), we name states by the same indexes we encountered previously when defining coloring table maps in coloring models. Henceforth, without loss of generality, we assume that if $\ca = \tuple{\castateset , \catransrel , \castate_0}$ denotes a CA, $\castateset \subseteq \coloringtableindexuniverse$ (Footnote \ref{footnote:index} still applies).

When we compose two connectors that have CA as their behavioral model, we can compute the CA of the composed connector by composing the CA of its constituents: the binary operator for CA composition takes the Cartesian product of the set of states of its arguments, designates the pair of their initial states as the initial state of the composed CA, and computes a new transition relation.%
\begin{definition}
	[Composition of CA \cite{baier06}]
	\label{def:cacomp}
	Let $\ca_1 = \tuple{\castateset_1 , \catransrel_1 , \castate_0^1}$ and $\ca_2 = \tuple{\castateset_2 , \catransrel_2 , \castate_0^2}$ be CA over $[\reonodeset_1 , \caconstraintset_1]$ and $[\reonodeset_2 , \caconstraintset_2]$. Their composition, denoted $\ca_1 \cacomp \ca_2$, is a CA over $[\reonodeset_1 \cup \reonodeset_2 , \caconstraintset_1 \wedge \caconstraintset_2]$%
	\footnote{%
		\label{footnote:wedge}%
		For notational convenience, we write $\caconstraintset_1 \wedge \caconstraintset_2$ for $\setleft \caconstraint_1 \wedge \caconstraint_2 \setbar \caconstraint_1 \in \caconstraintset_1 \mbox{ and } \caconstraint_2 \in \caconstraintset_2 \setright$.
	} defined as:
	\begin{center}$
		\ca_1 \cacomp \ca_2 = \tuple{\castateset_1 \times \castateset_2 , \catransrel , \tuple{\castate_0^1 , \castate_0^2}}
	$\end{center}
	\begin{center}$
		\mbox{with: } \catransrel = \left\{ \tuple{\tuple{\castate_1 , \castate_2} , \reofnodeset_1 \cup \reofnodeset_2 , \caconstraint_1 \wedge \caconstraint_2 , \tuple{\castate_1' , \castate_2'}} \left| \begin{array}{@{\;}l@{}}
			\tuple{\castate_1 , \reofnodeset_1 , \caconstraint_1 , \castate_1'} \in \catransrel_1 \mbox{ and } \tuple{\castate_2 , \reofnodeset_2 , \caconstraint_2 , \castate_2'} \in \catransrel_2
		\\	\mbox{and } \reofnodeset_1 \cap \reonodeset_2 = \reofnodeset_2 \cap \reonodeset_1
		\end{array} \right. \right\}
	$\end{center}
\end{definition}
\noindent The previous definition differs slightly from the one in \cite{baier06}: we do not \emph{implicitly} assume that all states have a silent $\tau$-transition (as in \cite{baier06}), but \emph{explicitly} include these transitions in our models (represented by transitions labeled with $\tuple{\emptyset , \top}$). Though essentially a matter of representation, it simplifies later proofs. 
\begin{definition}
	[Composition of $\ca$-connectors]
	\label{def:conncacomp}
	Let $\connca_1 = \tuple{\connstruct_1 , \ca_1}$ and $\connca_2 = \tuple{\connstruct_2 , \ca_2}$ be CA connectors over $[\reonodeset_1 , \caconstraintset_1]$ and $[\reonodeset_2 , \caconstraintset_2]$ such that $\connstruct_1 \connstructcomp \connstruct_2$ is defined. Their composition, denoted $\connca_1 \conncacomp \connca_2$, is an $\ca$-connector over $[\reonodeset_1 \cup \reonodeset_2 , \caconstraintset_1 \wedge \caconstraintset_2]$ defined as:
	\begin{center}$
		\connca_1 \conncacomp \connca_2 = \tuple {\connstruct_1 \connstructcomp \connstruct_2 , \ca_1 \cacomp \ca_2}
	$\end{center}
\end{definition}

\noindent To illustrate the previous definition, we depict the CA of \lossyfifo in Figure \ref{fig:pa:lossyfifo}.%
\footnote{%
	Similar to the coloring model of \lossyfifo in Figure \ref{fig:nextfun:lossyfifo}, its CA does not model its intended semantics: the transition $\tuple{\lossyfifoemptyindex , \{A\} , \top , \lossyfifoemptyindex}$ describes the inadmissible loss of data in the \textsc{empty} state (similar to $\coloring_{24}$ in Figure \ref{fig:nextfun:lossyfifo}).
}

\begin{figure}[t]
	\centering
	\scalebox{.8}{
	\begin{tabular}{|l|}
		\hline
	\\	\vspace{-2.25em}
	\\	\begin{tikzpicture}[baseline]
	\tikz[pin distance=.5cm];
	\node[state] 					(Q) [pin={[pin edge={black, thick, dotted}]below:\lossyfifoemptyindex}] {};
	\node[state, node distance=6cm]	(R) [right of=Q, pin={[pin edge={black, thick, dotted}]below:\lossyfifofullindex}] {};
	\path[->] (Q) edge [in=-165,	out=-15, thick]			node [below]	{\parbox{85.18126pt}{\footnotesize $\begin{array}{@{}c@{}}\{ A , M \} , \\ \#A = \#M \wedge \#M = \mbox{\upshape ``foo''}\end{array}$}} (R);
	\path[->] (Q) edge [in=-165,	out=165, thick, loop]	node [left]		{\footnotesize $\emptyset , \top$} (Q);
	\path[->] (Q) edge [in=135,		out=105, thick, loop]	node [above]	{\footnotesize $\{ A \} , \top$} (Q);
	\path[->] (R) edge [in=15,		out=165, thick]			node [above]	{\footnotesize $\{ B \} , \#B = \mbox{\upshape ``foo''}$} (Q);
	\path[->] (R) edge [in=60,		out=120, thick]			node [above]	{\footnotesize $\{ A , B \} , \#B = \mbox{\upshape ``foo''}$} (Q);
	\path[->] (R) edge [in=15,		out=-15, thick, loop]	node [right]	{\footnotesize $\emptyset , \top$} (R);
	\path[->] (R) edge [in=75,		out=45, thick, loop]	node [above]	{\footnotesize $\{ A \} , \top$} (R);
\end{tikzpicture}
	\\	\hline
	\end{tabular}
	}
	\caption{Constraint automaton of \lossyfifo with $\dataitemuniverse = \{ \mbox{\upshape ``foo''} \}$.}
	\label{fig:pa:lossyfifo}
\end{figure}

In the remainder, with a slight loss of generality, we consider only deterministic CA. We believe, however, that this limits the applicability of our results only marginally: in practice, non-deterministic CA occur rarely. In fact, even after almost a decade of research and literature on Reo, we have not encountered a (primitive) connector whose behavior one cannot describe concisely with a deterministic CA.%
\footnote{%
	Many non-deterministic connectors, in contrast, \emph{do} exist, but we can model their behavior with deterministic CA.
} As a final remark, we emphasize that our presentation of CA remains superficial: we covered only the essentials relevant to the rest of this paper. A more comprehensive overview appears in \cite{baier06}.

\newcommand{\ccoloring}				{\boldsymbol \coloring}
\newcommand{\ccoloringcomp}			{\boldsymbol \coloringcomp}
\newcommand{\ccoloringtable}		{\boldsymbol \coloringtable}
\newcommand{\ccoloringtablecomp}	{\boldsymbol \coloringtablecomp}
\newcommand{\ccoloringtablemap}		{\boldsymbol \coloringtablemap}
\newcommand{\ccoloringtablemapcomp}	{\boldsymbol \coloringtablemapcomp}
\newcommand{\cnextfun}			{\boldsymbol \nextfun}
\newcommand{\cnextfuncomp}		{\boldsymbol \nextfuncomp}
\newcommand{\cnextfuninit}		{\boldsymbol \nextfuninit}
\newcommand{\cnextfuninitcomp}	{\boldsymbol \nextfuninitcomp}

\newcommand{\connccol}			{\mathcal{C}^{\mbox{\upshape\tiny\bf\textsf{Col}}}}
\newcommand{\connccolcomp}		{\boldsymbol \conncolcomp}
\newcommand{\connccolinit}		{\connccol}
\newcommand{\connccolinitcomp}	{\boldsymbol \conncolinitcomp}

%
\section{Data-Aware Coloring Models}
\label{sect:constraintcc}

In this section, we make traditional coloring models \emph{data-aware} by extending them with constraints similar to those carried by transitions of constraint automata. We introduce this extension, because one of the transformation operators that we define later in this paper lacks a desirable property otherwise: it would map many-to-one instead of one-to-one. More precisely, the transformation from $\ca$-connectors to $\nextfuninit$-connectors would map \emph{different} $\ca$-connectors|i.e., those whose transitions carry different data constraints but equal firing sets|to the \emph{same} $\nextfuninit$-connector. Alternatively, to gain this desirable one-to-one property, we could have narrowed the scope of this paper to a special class of CA whose members abstract from data constraints: the transitions of these \emph{port automata} (PA) \cite{koehler09} carry only a firing set. We favor the extension of coloring models for generality (note that CA subsume PA).

We extend coloring models with data-awareness by associating each coloring with a data constraint from $\caconstraintuniverse$ (recall from Definitions \ref{def:caconstraintuniverse} and \ref{def:ca} that data constraints in CA come from the same universe). Such a \emph{constraint coloring} describes a computation step of a connector wherein (i) data flows through the nodes marked by the flow color and (ii) the data constraint holds. Below we give the formal definition. With respect to notation, we write the symbols that denote constituents of data-aware coloring models in \textbf{\emph{font}} (but use the same letters as for coloring models without constraints).
\begin{definition}
	[Constraint coloring]
	\label{def:ccoloring}
	Let $\reonodeset \subseteq \reonodeuniverse$ and $\caconstraintset \subseteq \caconstraintuniverse$. A constraint coloring $\ccoloring$ over $[\reonodeset , \caconstraintset]$ is a pair $\tuple{\coloring , \caconstraint}$ with $\coloring$ a coloring over $\reonodeset$ and $\caconstraint \in \caconstraintset$ a data constraint.
\end{definition}
\noindent Note that our definition does not exclude a constraint coloring $\ccoloring = \tuple{\coloring , \caconstraint}$ over $[\reonodeset , \caconstraintset]$ with inconsistent $\coloring$ and $\caconstraint$. For example, $\coloring$ may mark some node $\reonode \in \reonodeset$ with the no-flow color, while $\caconstraint \in \caconstraintset$ entails the flow of a data item $\dataitem \in \dataitemuniverse$ through $\reonode$. We do not forbid such constraint colorings, because they do not impair the models in which they appear: they merely describe behavior that cannot arise in practice.

Next, we incorporate data constraints in the definitions of the other constituents of ordinary coloring models (as presented in Section \ref{sect:prelim:cc}). This turns out straightforwardly. To summarize the upcoming definitions: (i) a \emph{constraint coloring table} is a set of constraint colorings, (ii) a \emph{constraint CTM} is a map from indexes to constraint coloring tables, (iii) a \emph{constraint next function} is map from [index, constraint coloring]-pairs to indexes, (iv) an \emph{initialized constraint next function} is a [constraint next function, index]-pair, and (v) an \emph{$\cnextfuninit$-connector} is a formal model of a connector that has an initialized constraint next function as its behavioral model.%
\begin{definition}
	[Constraint coloring table]
	\label{def:ccoloringtable}
	Let $\reonodeset \subseteq \reonodeuniverse$ and $\caconstraintset \subseteq \caconstraintuniverse$. A constraint coloring table $\ccoloringtable$ over $[\reonodeset , \caconstraintset]$ is a (sub)set (of) $\{ \reonodeset \rightarrow \coluniverse \} \times \caconstraintset$ of constraint colorings over $[\reonodeset , \caconstraintset]$.
\end{definition}
\begin{definition}
	[Constraint CTM]
	\label{def:ccoloringtablemap}
	Let $\reonodeset \subseteq \reonodeuniverse$, $\caconstraintset \subseteq \caconstraintuniverse$, and $\coloringtableindexset \subseteq \coloringtableindexuniverse$. A constraint CTM $\ccoloringtablemap$ over $[\reonodeset , \caconstraintset , \coloringtableindexset]$ is a map $\coloringtableindexset \rightarrow 2^{\{ \reonodeset \rightarrow \mbox{\scriptsize $\coluniverse$} \} \times \caconstraintset}$ from indexes to constraint coloring tables over $[\reonodeset , \caconstraintset]$.
\end{definition}
\begin{definition}
	[Constraint next function]
	\label{def:cnextfun}
	Let $\ccoloringtablemap$ be a constraint CTM over $[\reonodeset , \caconstraintset , \coloringtableindexset]$. A constraint next function $\cnextfun$ over $\ccoloringtablemap$ is a partial map $\coloringtableindexset \times ( \{ \reonodeset \rightarrow \coluniverse \} \times \caconstraintset ) \rightharpoonup \coloringtableindexset$ from [index, constraint coloring]-pairs to indexes such that $[\tuple{\coloringtableindex , \ccoloring} \mapsto \coloringtableindex'] \in \cnextfun$ iff $\ccoloring \in \ccoloringtablemap(\coloringtableindex)$.
\end{definition}
\begin{definition}
	[Initialized constraint next function]
	\label{def:cnextfuninit}
	Let $\ccoloringtablemap$ be a constraint CTM over $[\reonodeset , \caconstraintset , \coloringtableindexset]$. An initialized constraint next function $\cnextfuninit$ over $\ccoloringtablemap$ is a pair $\tuple{\cnextfun , \coloringtableindex_0}$ with $\cnextfun$ a constraint next function over $\ccoloringtablemap$ and $\coloringtableindex_0 \in \coloringtableindexset$.
\end{definition}
\begin{definition}
	[$\cnextfuninit$-connector]
	\label{def:connccolinit}
	Let $\ccoloringtablemap$ be a constraint CTM over $[\reonodeset , \caconstraintset , \coloringtableindexset]$. An $\cnextfuninit$-connector $\connccolinit$ over $[\reonodeset , \ccoloringtablemap]$ is a pair $\tuple{\connstruct , \cnextfuninit}$ with $\connstruct = \tuple{\reonodeset , \iomap}$ a connector structure and $\cnextfuninit$ an initialized constraint next function over $\ccoloringtablemap$. 
\end{definition}

To illustrate the previous definitions, Figure \ref{fig:cnextfun:prim} shows the constraint CTMs of \sync, \lossysync, and \fifo. The colorings $c_i$ with $i \in \{1,2,3,4\}$ refer to the colorings in Figure \ref{fig:nextfun:prim}. For instance, constraint coloring $\tuple{\coloring_1 , \#A = \#B}$ in the constraint CTM of \sync (respectively \lossysync) describes the behavior alternative of \sync (respectively \lossysync) wherein the same data item flows through $A$ and $B$. Constraint coloring $\tuple{\coloring_4 , \top}$, present in all constraint CTMs, describes the behavior alternative wherein a connector idles. Due to the constraint $\top$, this may always happen. Similarly, \lossysync can always behave as described by $\tuple{\coloring_2 , \top}$, but whereas $\tuple{\coloring_4 , \top}$ entails no flow at all, $\tuple{\coloring_2, \top}$ entails flow through $A$ and no flow through $B$. Here, $\top$ specifies that we do not care about which data item flows through $A$. With respect to \fifo, for simplicity of the example, we assume the universe of data items a singleton similar to Section \ref{sect:prelim:ca}. In general, the constraint CTM of \fifo contains a distinct constraint coloring table for each data item in $\dataitemuniverse$ that may occupy the buffer.

\begin{figure}[t]
	\centering
	\scalebox{.8}{
	\begin{tabular}{|c|c|c|}
		\hline &&
	\\	\vspace{-1.75em} &&
	\\	\tablefill{3em} $\begin{array}[t]{@{}l@{\:}c@{\:}l@{}}
			\ccoloringtablemap & = & \left\{ \syncindex \mapsto \left\{\begin{array}{@{}l@{}}
				\tuple{\coloring_1 , \; \#A = \#B} \; ,
			\\	\tuple{\coloring_4 , \; \top}
			\end{array} \right\} \right\}
		\end{array}$
	&	$\begin{array}[t]{@{}l@{\:}c@{\:}l@{}}
			\ccoloringtablemap & = & \left\{ \lossysyncindex \mapsto \left\{\begin{array}{@{}l@{}}
				\tuple{\coloring_1 , \; \#A = \#B} \; ,
			\\	\tuple{\coloring_2 , \; \top} \; ,
			\\	\tuple{\coloring_4 , \; \top}
			\end{array} \right\} \right\}
		\end{array}$
	&	$\begin{array}[t]{@{}l@{\:}c@{\:}l@{}}
			\ccoloringtablemap & = & \left\{ \begin{array}{@{}l@{}}
				\fifoemptyindex \mapsto \left\{\begin{array}{@{}l@{}}
				\tuple{\coloring_2 , \; \#A = \mbox{\upshape ``foo''}} \; ,
			\\	\tuple{\coloring_4 , \; \top}
			\end{array} \right\} \; ,
			\\	\vspace{-.75em}
			\\	\fifofullindex \mapsto \left\{\begin{array}{@{}l@{}}
				\tuple{\coloring_3 , \; \#B = \mbox{\upshape ``foo''}} \; ,
			\\	\tuple{\coloring_4 , \; \top}
			\end{array} \right\}
			\end{array} \right\}
		\end{array}$
	\\	\hline
	\end{tabular}
	}
	\caption{Constraint CTMs of \sync, \lossysync, and \fifo with $\dataitemuniverse = \{\mbox{\upshape ``foo''}\}$.}
	\label{fig:cnextfun:prim}
\end{figure}

Finally, we must update the composition operators for coloring models to incorporate data constraints. For brevity, we give these definitions only for constraint colorings and constraint coloring tables. The composition operators for constraint CTMs (symbol: $\ccoloringtablemapcomp$\label{def:ccoloringtablemapcomp:info}), (initialized) constraint next functions (symbol: $\cnextfuncomp$\label{def:cnextfuncomp:info}\label{def:cnextfuninitcomp:info}), and $\cnextfuninit$-connectors (symbol: $\connccolinitcomp$\label{def:connccolinitcomp:info}) resemble their respective composition operators in Section \ref{sect:prelim:cc}: essentially, it suffices to replace $\coloringtablemap_1$, $\coloringtablemap_2$, $\nextfun_1$, $\nextfun_2$, $\nextfuninit_1$, and $\nextfuninit_2$ in Definitions \ref{def:coloringtablemapcomp}--\ref{def:conncolinitcomp} with their \textbf{\emph{font}} versions $\ccoloringtablemap_1$, $\ccoloringtablemap_2$, $\cnextfun_1$, $\cnextfun_2$, $\cnextfuninit_1$, and $\cnextfuninit_2$. We require only these minor updates, because our extension of coloring models with data constraints affects only the definition of colorings directly. For completeness, in Appendix \ref{appx}, we give the definitions of those composition operators that we skip below.%
\begin{definition}
	[Composition of constraint colorings]
	\label{def:ccoloringcomp}
	Let $\ccoloring_1 = \tuple{\coloring_1 , \caconstraint_1}$ and $\ccoloring_2 = \tuple{\coloring_2 , \caconstraint_2}$ be colorings over $[\reonodeset_1 , \caconstraintset_1]$ and $[\reonodeset_2 , \caconstraintset_2]$ such that $\coloring_1(\reonode) = \coloring_2(\reonode)$ for all $\reonode \in \reonodeset_1 \cap \reonodeset_2$. Their composition, denoted by $\ccoloring_1 \ccoloringcomp \ccoloring_2$, is a constraint coloring over $[\reonodeset_1 \cup \reonodeset_2 , \caconstraintset_1 \wedge \caconstraintset_2]^{\ref{footnote:wedge}}$ defined as:
	\begin{center}$
		\ccoloring_1 \ccoloringcomp \ccoloring_2 = \tuple{\coloring_1 \coloringcomp \coloring_2 , \caconstraint_1 \wedge \caconstraint_2}
	$\end{center}
\end{definition}
\begin{definition}
	[Composition of constraint coloring tables]
	\label{def:ccoloringtablecomp}
	Let $\ccoloringtable_1$ and $\ccoloringtable_2$ be constraint coloring tables over $[\reonodeset_1 , \caconstraintset_1]$ and $[\reonodeset_2 , \caconstraintset_2]$. Their composition, denoted by $\ccoloringtable_1 \ccoloringtablecomp \ccoloringtable_2$, is a constraint coloring table over $[\reonodeset_1 \cup \reonodeset_2 , \caconstraintset_1 \wedge \caconstraintset_2]$ defined as:
	\begin{center}$
		\ccoloringtable_1 \ccoloringtablecomp \ccoloringtable_2 = \setleft \ccoloring_1 \ccoloringcomp \ccoloring_2 \setbar \ccoloring_1 = \tuple{\coloring_1 , \caconstraint_1} \in \ccoloringtable_1 \mbox{ and } \ccoloring_2 = \tuple{\coloring_2 , \caconstraint_2} \in \ccoloringtable_2 \mbox{ and } \coloring_1(\reonode) = \coloring_2(\reonode) \mbox{ for all } \reonode \in \reonodeset_1 \cap \reonodeset_2 \setright
	$\end{center}
\end{definition}
\noindent Note that the composition operator for CA in Definition \ref{def:cacomp} computes the label of a composite transition by taking the conjunction of two data constraints, similar to how we handle the combination of data constraints in Definition \ref{def:ccoloringcomp}. The lemmas that we formulate and prove in the subsequent sections establish the appropriateness of taking the conjunction of data constraints in the context of coloring models. To illustrate the previous definitions, Figure \ref{fig:cnextfun:lossyfifo} shows the constraint CTM of \lossyfifo. The colorings $\coloring_i$ with $i \in \{12 , 24 , 43 , 44\}$ refer to the colorings in Figure \ref{fig:nextfun:lossyfifo}.

\begin{figure}[t]
	\centering
	\scalebox{.8}{
	\begin{tabular}{|c|}
		\hline
 	\\	\vspace{-1.75em}
	\\	\tablefill{2.75em} $\begin{array}[t]{@{}l@{\:}c@{\:}l@{}}
			\ccoloringtablemap & = & \left\{ \begin{array}{@{}l@{}}
				\lossyfifoemptyindex \mapsto \left\{\begin{array}{@{}l@{}}
					\tuple{\coloring_{12} , \; \#A = \#M \wedge \#M = \mbox{\upshape ``foo''}} \; ,
				\\	\tuple{\coloring_{24} , \; \top} \; ,
				\\	\tuple{\coloring_{44} , \; \top}
				\end{array} \right\} \; , \; \lossyfifofullindex \mapsto \left\{\begin{array}{@{}l@{}}
					\tuple{\coloring_{12} , \; \#B = \mbox{\upshape ``foo''}} \; ,
				\\	\tuple{\coloring_{24} , \; \top} \; ,
				\\	\tuple{\coloring_{43} , \; \#B = \mbox{\upshape ``foo''}} \; ,
				\\	\tuple{\coloring_{44} , \; \top}
				\end{array} \right\}
			\end{array} \right\}
		\end{array}$
	\\	\hline
	\end{tabular}
	}
	\caption{Constraint CTM of \lossyfifo with $\dataitemuniverse = \{\mbox{\upshape ``foo''}\}$. We abbreviate $\tuple{\lossysyncindex , \fifoemptyindex}$ by $\lossyfifoemptyindex$ and $\tuple{\lossysyncindex , \fifofullindex}$ by $\lossyfifofullindex$.}
	\label{fig:cnextfun:lossyfifo}
\end{figure}

Recently, in \cite{proenca11}, Proen\c{c}a uses pairs of colorings and \emph{node-to-data functions} as transition labels of his \emph{behavioral automata} to account for the transfer of data that takes place through data-flows described by those colorings. This suggests using [coloring, node-to-data function]-pairs as a data-aware coloring model. However, our constraint-based extension offers a more concise formalization. For instance, in Proen\c{c}a's model, to give the semantics of a \sync primitive, one must include a separate [coloring, node-to-data function]-pair in its coloring table for each data item in $\dataitemuniverse$. With our constraint-based extension, in contrast, we capture this with a single constraint coloring as shown in Figure \ref{fig:cnextfun:prim}.

\newcommand{\toca}{\mathbb{L}}

%
\section{From \texorpdfstring{$\cnextfuninit$}{epsilon}-Connectors to  \texorpdfstring{$\ca$}{alpha}-Connectors}
\label{sect:cctoca}

In this section, we present a unary operator, denoted by $\toca$, which takes as argument an $\cnextfuninit$-connector and produces an \emph{equivalent} $\ca$-connector; shortly, we elaborate on the meaning of ``equivalance'' in this context. We call our process of transforming an $\cnextfuninit$-connector to an $\ca$-connector the \emph{$\toca$-transformation}. By defining the $\toca$-transformation for \emph{any} $\cnextfuninit$-connector, it follows that the class of connectors that we can model as $\ca$-connector \emph{includes} those that we can model as $\cnextfuninit$-connector|i.e., constraint automata are \emph{at least} as expressive as data-aware coloring models.

The $\toca$-operator works as follows; suppose we wish to transform an $\cnextfuninit$-connector $\connccolinit = \tuple{\connstruct , \cnextfuninit}$ over $[\reonodeset , \ccoloringtablemap]$ with $\ccoloringtablemap$ a constraint CTM over $[\reonodeset , \caconstraintset , \coloringtableindexset]$. Whereas the connector structure $\connstruct$ does not incur any change (because $\toca$ alters only the behavioral model), from the initialized constraint next function $\cnextfuninit = \tuple{\cnextfun , \coloringtableindex_0}$, the $\toca$-operator derives a constraint automaton. First, $\toca$ instantiates the set of states of this derived CA with the set of indexes $\coloringtableindexset$: this seems reasonable as $\coloringtableindexset$ denotes the set of indexes that represent the states of the connector that $\connccolinit$ models. Next, $\toca$ constructs a transition relation $\catransrel$ based on the mappings in $\cnextfun$: for each $[\tuple{\coloringtableindex , \tuple{\coloring , \caconstraint}} \mapsto \coloringtableindex'] \in \cnextfun$, the $\toca$-operator creates a transition from state $\coloringtableindex$ to state $\coloringtableindex'$, labeled with $\caconstraint$ as its data constraint and with the set of nodes to which $\coloring$ assigns the flow color as its firing set. Finally, $\coloringtableindex_0$ becomes the initial state of the new CA.%
\begin{definition}
	[$\toca$ for $\cnextfuninit$-connectors]
	\label{def:toca}
	Let $\connccolinit = \tuple{\connstruct , \cnextfuninit}$ be an $\cnextfuninit$-connector over $[\reonodeset , \ccoloringtablemap]$ with $\cnextfuninit = \tuple{\cnextfun , \coloringtableindex_0}$ and $\ccoloringtablemap$ a constraint CTM over $[\reonodeset , \caconstraintset , \coloringtableindexset]$. The $\toca$-transformation of $\connccolinit$, denoted by $\toca(\connccolinit)$, is defined as:
	\begin{center}$
		\begin{array}{@{}r@{\;}l@{}}
				\multicolumn{2}{c}{\toca(\connccolinit) = \tuple{\connstruct , \toca(\cnextfuninit)}}
			\\	\mbox{with:}
			&	\toca(\cnextfuninit) = \tuple{\coloringtableindexset , \catransrel , \coloringtableindex_0}
			\\	\mbox{and:}
			&	\catransrel = \setleft \tuple{\coloringtableindex , \reofnodeset , \caconstraint , \cnextfun(\coloringtableindex , \ccoloring)} \setbar \coloringtableindex \in \coloringtableindexset \mbox{ and } \ccoloring = \tuple{\coloring , \caconstraint} \in \ccoloringtablemap(\coloringtableindex) \mbox{ and } \reofnodeset = \setleft \reonode \in \reonodeset \setbar \coloring(\reonode) = \colsymflow \setright \setright
		\end{array}
	$\end{center}
\end{definition}
\noindent The following proposition states that the application of $\toca$ to an $\nextfun$-connector yields an $\ca$-connector.
\begin{proposition}
	\label{prop:wf:toca}
	Let $\connccolinit$ be an $\cnextfuninit$-connector over $[\reonodeset , \ccoloringtablemap]$ with $\ccoloringtablemap$ defined over $[\reonodeset , \caconstraintset , \coloringtableindexset]$. Then, $\toca(\connccolinit)$ is an $\ca$-connector over $[\reonodeset , \caconstraintset]$.
	
	\vspace{.5em}
	\barproof{
		Let $\connccolinit = \tuple{\connstruct , \cnextfuninit}$ with $\cnextfuninit = \tuple{\cnextfun , \coloringtableindex_0}$. Then, by Definition \ref{def:toca}, $\toca(\connccolinit) = \tuple{\connstruct , \toca(\cnextfuninit)} = \tuple{\connstruct , \tuple{\coloringtableindexset , \catransrel , \coloringtableindex_0}}$. By Definition \ref{def:connca}, we must show that $\tuple{\coloringtableindexset , \catransrel , \coloringtableindex_0}$ is a CA over $[\reonodeset , \caconstraintset]$. To demonstrate this, by Definition \ref{def:ca}, we must show that $\catransrel \subseteq \coloringtableindexset \times 2^{\reonodeset} \times \caconstraintset \times \coloringtableindexset$. Because, by the premise, $\connccolinit$ is an $\cnextfuninit$-connector over $[\reonodeset , \ccoloringtablemap]$, by Definition \ref{def:connccolinit}, $\cnextfuninit$ is an initialized constraint next function over $\ccoloringtablemap$, hence, by Definition \ref{def:cnextfuninit}, the co-domain of $\cnextfun$ is $\coloringtableindexset$. Also, by Definition \ref{def:toca}, for all $\tuple{\coloringtableindex , \reofnodeset , \caconstraint , \cnextfun(\coloringtableindex , \ccoloring)} \in \catransrel$ with $\ccoloring = \tuple{\coloring , \caconstraint}$, it holds that $\coloringtableindex \in \coloringtableindexset$, $\reofnodeset \subseteq \reonodeset$, and $\caconstraint \in \caconstraintset$ (this latter follows from Definition \ref{def:ccoloring}).
	}
	\vspace{.25em}
\end{proposition}

\noindent In the rest of this section, we prove the equivalence between an $\cnextfuninit$-connector $\connccolinit$ and the $\ca$-connector that results from applying $\toca$ to $\connccolinit$. Additionally, we prove the distributivity of $\toca$ over composition.

\newcommand{\carun}			{r}
\newcommand{\carunset}		{\boldsymbol{\rho}}

\newcommand{\painting}		{p}
\newcommand{\paintingset}	{\boldsymbol{\pi}}
\newcommand{\prequiv}		{\asymp}

\newcommand{\tocabisim}		{\sim}
\newcommand{\tocabisimrel}	{\mathcal{R}}

\subsection{Correctness of \texorpdfstring{$\toca$}{L}}
\label{sect:cctoca:corr}

In this subsection, we prove the \emph{correctness} of $\toca$: we consider $\toca$ correct if its application to an $\cnextfuninit$-connector yields an \emph{equivalent} $\ca$-connector. We call an $\cnextfuninit$-connector and an $\ca$-connector equivalent if there exists a \emph{bi-simulation relation} that relates these two connector models. Informally, an $\ca$-connector $\connca$ is bi-similar to an $\cnextfuninit$-connector $\connccolinit$ if, for each mapping in the constraint next function of $\connccolinit$, there exists a \emph{corresponding} transition in the CA of $\connca$|i.e., a transition that describes the same behavior in terms of the nodes that fire, the data items that flow, and the change of state|and vice versa.%
\begin{definition}
	[Bi-simulation]
	\label{def:tocabisim}
	Let $\connca = \tuple{\connstruct , \ca}$ with $\ca = \tuple{\castateset , \catransrel , \castate_0}$ be an $\ca$-connector over $[\reonodeset , \caconstraintset]$ and $\connccolinit = \tuple{\connstruct , \cnextfuninit}$ with $\cnextfuninit = \tuple{\cnextfun , \coloringtableindex_0}$ an $\cnextfuninit$-connector over $[\reonodeset , \ccoloringtablemap]$ with $\ccoloringtablemap$ defined over $[\reonodeset , \caconstraintset , \coloringtableindexset]$. $\connca$ and $\connccolinit$ are bi-similar, denoted as $\connca \tocabisim \connccolinit$, if there exists a relation $\tocabisimrel \subseteq \castateset \times \coloringtableindexset$ such that $\tuple{\castate_0 , \coloringtableindex_0} \in \tocabisimrel$ and for all $\tuple{\castate , \coloringtableindex} \in \tocabisimrel$:
	\begin{center}
		\begin{minipage}{.45\textwidth}
			\textsc{(i)} If $\tuple{\castate , \reofnodeset , \caconstraint , \castate'} \in \catransrel$ then there exists a $\coloringtableindex' \in \coloringtableindexset$ such that:
			\begin{itemize}
				\setlength{\itemsep}{0cm}
  				\setlength{\parskip}{0cm}
				\item $[\tuple{\coloringtableindex , \ccoloring} \mapsto \coloringtableindex'] \in \cnextfun$ with $\ccoloring = \tuple{\coloring , \caconstraint}$;
				\item $\tuple{\castate' , \coloringtableindex'} \in \tocabisimrel$;
				\item $\reofnodeset = \setleft \reonode \in \reonodeset \setbar \coloring(\reonode) = \colsymflow \setright$.
			\end{itemize}
		\end{minipage}%
		\hspace{.05\textwidth}%
		\begin{minipage}{.45\textwidth}
			\textsc{(ii)} If $[\tuple{\coloringtableindex , \ccoloring} \mapsto \coloringtableindex'] \in \cnextfun$ with $\ccoloring = \tuple{\coloring, \caconstraint}$ then there exists a $\castate' \in \castateset$ such that:
			\begin{itemize}
				\setlength{\itemsep}{0cm}
  				\setlength{\parskip}{0cm}
				\item $\tuple{\castate , \reofnodeset , \caconstraint , \castate'} \in \catransrel$;
				\item $\tuple{\castate' , \coloringtableindex'} \in \tocabisimrel$;
				\item $\reofnodeset = \setleft \reonode \in \reonodeset \setbar \coloring(\reonode) = \colsymflow \setright$.
			\end{itemize}
		\end{minipage}
	\end{center}
	In that case, $\tocabisimrel$ is called a bi-simulation relation.
\end{definition}
\noindent Next, we formulate and prove Lemma \ref{lemma:corr:bisim:toca}, which states the bi-similarity between an $\cnextfuninit$-connector $\connccolinit$ and its $\toca$-transformation $\toca(\connccolinit)$. In our proof, we choose the diagonal relation on the set of indexes as a candidate bi-simulation relation.
\begin{lemma}
	\label{lemma:corr:bisim:toca}
	Let $\connccolinit$ be an $\cnextfuninit$-connector. Then, $\toca(\connccolinit) \tocabisim \connccolinit$.
	
	\vspace{.5em}
	\barproof{%
		Let $\connccolinit = \tuple{\connstruct , \cnextfuninit}$ with $\cnextfuninit = \tuple{\cnextfun , \coloringtableindex_0}$ be defined over $[\reonodeset , \ccoloringtablemap]$ with $\ccoloringtablemap$ defined over $[\reonodeset , \caconstraintset , \coloringtableindexset]$. Additionally, let $\toca(\connccolinit) = \tuple{\connstruct , \toca(\cnextfuninit)}$ with $\toca(\cnextfuninit) = \tuple{\coloringtableindexset , \catransrel , \coloringtableindex_0}$. We show that $\tocabisimrel = \setleft \tuple{\coloringtableindex , \coloringtableindex} \setbar \coloringtableindex \in \coloringtableindexset \setright$ is a bi-simulation relation by demonstrating that it satisfies \textsc{(i)} and \textsc{(ii)} of Definition \ref{def:tocabisim}. Let $\tuple{\coloringtableindex , \coloringtableindex} \in \tocabisimrel$.
		\begin{description}
			\item[\rm\textsc{(i)}] Suppose $\tuple{\coloringtableindex , \reofnodeset , \caconstraint , \coloringtableindex'} \in \catransrel$. Then, by Definition \ref{def:toca} of $\toca$, there exists a $\ccoloring = \tuple{\coloring , \caconstraint} \in \ccoloringtablemap(\coloringtableindex)$ such that $\coloringtableindex' = \cnextfun(\coloringtableindex , \ccoloring)$. Hence, $[\tuple{\coloringtableindex , \ccoloring} \mapsto \coloringtableindex'] \in \cnextfun$. Also, by the definition of $\tocabisimrel$, $\tuple{\coloringtableindex' , \coloringtableindex'} \in \tocabisimrel$. Finally, by Definition \ref{def:toca} of $\toca$, $\reofnodeset = \setleft \reonode \in \reonodeset \setbar \coloring(\reonode) = \colsymflow \setright$. Therefore, $\tocabisimrel$ satisfies \textsc{(i)}.
			\item[\rm\textsc{(ii)}] Suppose $[\tuple{\coloringtableindex , \ccoloring} \mapsto \coloringtableindex'] \in \cnextfun$ with $\ccoloring = \tuple{\coloring, \caconstraint}$. Then, by Definition \ref{def:cnextfun} of $\cnextfun$, it holds that $\coloringtableindex \in \coloringtableindexset$ and $\ccoloring \in \ccoloringtablemap(\coloringtableindex)$, hence by Definition \ref{def:toca} of $\toca$, $\tuple{\coloringtableindex , \reofnodeset , \caconstraint , \coloringtableindex'} \in \catransrel$ with $\reofnodeset = \setleft \reonode \in \reonodeset \setbar \coloring(\reonode) = \colsymflow \setright$. Also, by the definition of $\tocabisimrel$, $\tuple{\coloringtableindex' , \coloringtableindex'} \in \tocabisimrel$. Therefore, $\tocabisimrel$ satisfies \textsc{(ii)}.
		\end{description}
		Thus, $\tocabisimrel$ satisfies \textsc{(i)} and \textsc{(ii)}. Also, because $\coloringtableindex_0 \in \coloringtableindexset$ by Definition \ref{def:cnextfuninit}, $\tuple{\coloringtableindex_0 , \coloringtableindex_0} \in \tocabisimrel$. Hence, $\tocabisimrel$ is a bi-simulation relation. Therefore, $\toca(\connccolinit) \tocabisim \connccolinit$.
	}
\end{lemma}

\subsection{Distributivity of \texorpdfstring{$\toca$}{L}}
\label{sect:cctoca:dist}

We end this section with the distributivity (compositionality) lemma of $\toca$. Informally, it states that it does not matter whether we first compose $\cnextfuninit$-connectors $\connccolinit_1$ and $\connccolinit_2$ and then apply $\toca$ to the composition or first apply $\toca$ to $\connccolinit_1$ and $\connccolinit_2$ and then compose the transformations; the resulting $\ca$-connectors equal each other. The relevance of this result lies in the potential reduction in the amount of overhead that it allows for when applying the $\toca$-operator in practice. This works as follows. There exist tools for Reo that operate on constraint automata and that have built-in functionality for their composition. By the distributivity lemma of $\toca$, to use such a tool, we need to transform Reo's common primitives only once, store these in a library, and use this library together with the built-in functionality for composition to construct the CA of composed connectors (on which the tool subsequently operates). Thus, the overhead of this approach remains constant. In contrast, the overhead of the alternative|i.e., first composing coloring models and then transforming the resulting composition using $\toca$|grows linear in the number of composed connectors one wishes to apply the tool on.%
	\footnote{%
		In the current exposition, we assume composing two $\cnextfuninit$-connectors and two $\ca$-connectors have equal costs. In Section \ref{sect:appl}, we argue for the merits of our approach when the cost of coloring model composition differs from the cost of CA composition.
} In Section \ref{sect:appl}, we illustrate the foregoing with a concrete example; here, we proceed with the lemma and our proof, which, although rather technical, essentially consists of a series of straightforward applications of definitions that allow us to rewrite the transition relations of the composed automata. 
\begin{lemma}
	\label{lemma:dist:toca}
	Let $\connccolinit_1$ and $\connccolinit_2$ be $\cnextfuninit$-connectors. Then, $\toca(\connccolinit_1) \conncacomp \toca(\connccolinit_2) = \toca(\connccolinit_1 \connccolinitcomp \connccolinit_2)$.
	
	\vspace{.5em}
	\barproof{%
		Let $\connccolinit_1 = \tuple{\connstruct_1 , \cnextfuninit_1}$ and $\connccolinit_2 = \tuple{\connstruct_2 , \cnextfuninit_2}$ with $\cnextfuninit_1 = \tuple{\cnextfun_1 , \coloringtableindex_0^1}$ and $\cnextfuninit_2 = \tuple{\cnextfun_2 , \coloringtableindex_0^2}$ be defined over $[\reonodeset_1 , \ccoloringtablemap_1]$ and $[\reonodeset_2 , \ccoloringtablemap_2]$. Applying Definition \ref{def:toca} of $\toca$, Definition \ref{def:conncacomp} of $\conncacomp$, and Definition \ref{def:connccolinitcomp} of $\connccolinitcomp$ (informally on page \pageref{def:connccolinitcomp:info}) yields:
		\begin{center}$
			\tuple{\connstruct_1 \connstructcomp \connstruct_2 , \toca(\cnextfuninit_1) \cacomp \toca(\cnextfuninit_2)} = \tuple{\connstruct_1 \connstructcomp \connstruct_2 , \toca(\cnextfuninit_1 \cnextfuncomp \cnextfuninit_2)}
		$\end{center}
		We focus on proving $\toca(\cnextfuninit_1) \cacomp \toca(\cnextfuninit_2) = \toca(\cnextfuninit_1 \cnextfuncomp \cnextfuninit_2)$. Applying Definition \ref{def:toca} of $\toca$ to the left-hand side (LHS) and Definition \ref{def:cnextfuninitcomp} of $\cnextfuninitcomp$ (informally on page \pageref{def:cnextfuninitcomp:info}) to the right-hand side (RHS) yields:
		\begin{center}$
			\begin{array}{@{}r@{\;}l@{}}
				\multicolumn{2}{c}{%
					\begin{array}{@{}c@{}}
						\tuple{\coloringtableindexset_1 , \catransrel_1 , \coloringtableindex_0^1}
					\\	\cacomp
					\\	\tuple{\coloringtableindexset_2 , \catransrel_2 , \coloringtableindex_0^2}
					\end{array} = \toca \left( \left\langle \left\{ \left. \begin{array}{@{}c@{\;}}
						\tuple{\coloringtableindex_1 , \coloringtableindex_2} , \ccoloring_1 \ccoloringcomp \ccoloring_2 
					\\	\rotatebox[origin=c]{270}{$\mapsto$}
					\\	\tuple{\cnextfun_1(\coloringtableindex_1 , \ccoloring_1) , \cnextfun_2(\coloringtableindex_2 , \ccoloring_2)}
					\end{array} \right| \begin{array}{@{\;}c@{}}
						\tuple{\coloringtableindex_1 , \coloringtableindex_2} \in \coloringtableindexset_1 \times \coloringtableindexset_2 
					\\	\mbox{and}
					\\	\ccoloring_1 \ccoloringcomp \ccoloring_2 \in (\ccoloringtablemap_1 \ccoloringtablemapcomp \ccoloringtablemap_2)(\tuple{\coloringtableindex_1 , \coloringtableindex_2})
					\end{array} \right\} , \tuple{\coloringtableindex_0^1, \coloringtableindex_0^2} \right\rangle \right)
				}
			\\	\vspace{-.75em}
			\\	\mbox{with:}
			&	\catransrel_1 = \left\{ \tuple{\coloringtableindex_1 , \reofnodeset_1 , \caconstraint_1 , \cnextfun_1(\coloringtableindex_1 , \ccoloring_1)} \left| \begin{array}{@{\;}l@{}}
					\coloringtableindex_1 \in \coloringtableindexset_1 \mbox{ and } \ccoloring_1 = \tuple{\coloring_1 , \caconstraint_1} \in \ccoloringtablemap_1(\coloringtableindex_1) \mbox{ and}
				\\	\reofnodeset_1 = \setleft \reonode \in \reonodeset_1 \setbar \coloring_1(\reonode) = \colsymflow \setright
				\end{array} \right. \right\}
			\\	\vspace{-.75em}
			\\	\mbox{and:}
			&	\catransrel_2 = \left\{ \tuple{\coloringtableindex_2 , \reofnodeset_2 , \caconstraint_2 , \cnextfun_2(\coloringtableindex_2 , \ccoloring_2)} \left| \begin{array}{@{\;}l@{}}
					\coloringtableindex_2 \in \coloringtableindexset_2 \mbox{ and } \ccoloring_2 = \tuple{\coloring_2 , \caconstraint_2} \in \ccoloringtablemap_1(\coloringtableindex_2) \mbox{ and}
				\\	\reofnodeset_2 = \setleft \reonode \in \reonodeset_2 \setbar \coloring_2(\reonode) = \colsymflow \setright
				\end{array} \right. \right\}
			\end{array}
		$\end{center}
		Applying Definition \ref{def:cacomp} of $\cacomp$ to the LHS, and Definition \ref{def:toca} of $\toca$ to the RHS yields:
		\begin{center}$
			\begin{array}{@{}r@{\;}l@{}}
				\multicolumn{2}{c}{%
					\tuple{\coloringtableindexset_1 \times \coloringtableindexset_2 , \catransrel , \tuple{\coloringtableindex_0^1 , \coloringtableindex_0^2}} = \tuple{\coloringtableindexset_1 \times \coloringtableindexset_2 , \catransrel' , \tuple{\coloringtableindex_0^1 , \coloringtableindex_0^2}}
				}
			\\	\vspace{-.75em}
			\\	\mbox{with:}
			&	\catransrel = \left\{ \tuple{\tuple{\coloringtableindex_1 , \coloringtableindex_2} , \reofnodeset_1 \cup \reofnodeset_2 , \caconstraint_1 \wedge \caconstraint_2 , \tuple{\coloringtableindex_1' , \coloringtableindex_2'}} \left| \begin{array}{@{\;}l@{}}
					\tuple{\coloringtableindex_1 , \reofnodeset_1 , \caconstraint_1 , \coloringtableindex_1'} \in \catransrel_1 \mbox{ and } \tuple{\coloringtableindex_2 , \reofnodeset_2 , \caconstraint_2 , \coloringtableindex_2'} \in \catransrel_2 
				\\	\mbox{and } \reofnodeset_1 \cap \reonodeset_2 = \reofnodeset_2 \cap \reonodeset_1
				\end{array} \right. \right\}
			\\	\vspace{-.75em}
			\\	\mbox{and:}
			&	\catransrel' = \left\{ \tuple{\coloringtableindex , \reofnodeset , \caconstraint , (\cnextfun_1 \nextfuncomp \cnextfun_2)(\coloringtableindex , \ccoloring)} \left| \begin{array}{@{\;}l@{}}
					\coloringtableindex \in \coloringtableindexset_1 \times \coloringtableindexset_2 \mbox{ and } \ccoloring = \tuple{\coloring , \caconstraint} \in (\ccoloringtablemap_1 \ccoloringtablemapcomp \ccoloringtablemap_2)(\coloringtableindex) 
				\\	\mbox{and } \reofnodeset = \setleft \reonode \in \reonodeset_1 \cup \reonodeset_2 \setbar \coloring(\reonode) = \colsymflow \setright
				\end{array} \right. \right\}
			\end{array}
		$\end{center}
		What remains to be shown is $\catransrel = \catransrel'$. This follows from Figure \ref{fig:proof:lemma:dist:toca}.
	}
	
	\vspace{.5em}
	
	\begin{figure}[t]
		\centering
 		\fbox{\scalebox{.75}{$
			\begin{array}{@{}c@{\;}l@{}}
					& \catransrel
			\\	=	& \pcomment{By the definition of $\catransrel$ in Lemma \ref{lemma:dist:toca}}
			\\		& \setleft \tuple{\tuple{\coloringtableindex_1 , \coloringtableindex_2} , \reofnodeset_1 \cup \reofnodeset_2 , \caconstraint_1 \wedge \caconstraint_2 , \tuple{\coloringtableindex_1' , \coloringtableindex_2'}} \setbar
							\tuple{\coloringtableindex_1 , \reofnodeset_1 , \caconstraint_1 , \coloringtableindex_1'} \in \catransrel_1 \mbox{ and } \tuple{\coloringtableindex_2 , \reofnodeset_2 , \caconstraint_2 , \coloringtableindex_2'} \in \catransrel_2 \mbox{ and } \reofnodeset_1 \cap \reonodeset_2 = \reofnodeset_2 \cap \reonodeset_1
						\setright
			\\	=	& \pcomment{By the definitions of $\catransrel_1$ and $\catransrel_2$ in Lemma \ref{lemma:dist:toca}, and by introducing $\coloringtableindex = \tuple{\coloringtableindex_1 , \coloringtableindex_2}$}
			\\		& \left\{ \tuple{\coloringtableindex , \reofnodeset_1 \cup \reofnodeset_2 , \caconstraint_1 \wedge \caconstraint_2 , \tuple{\coloringtableindex_1' , \coloringtableindex_2'}} \left| \begin{array}{l}
							\coloringtableindex_1 \in \coloringtableindexset_1 \mbox{ and } \ccoloring_1 = \tuple{\coloring_1 , \caconstraint_1} \in \ccoloringtablemap_1(\coloringtableindex_1) \mbox{ and } \coloringtableindex_1' = \cnextfun_1(\coloringtableindex_1 , \ccoloring_1)
							\mbox{ and } \reofnodeset_1 = \setleft \reonode \in \reonodeset_1 \setbar \coloring_1(\reonode) = \colsymflow \setright \mbox{ and}
						\\	\coloringtableindex_2 \in \coloringtableindexset_2 \mbox{ and } \ccoloring_2 = \tuple{\coloring_2 , \caconstraint_2} \in \ccoloringtablemap_2(\coloringtableindex_2) \mbox{ and } \coloringtableindex_2' = \cnextfun_2(\coloringtableindex_2 , \ccoloring_2)
							\mbox{ and } \reofnodeset_2 = \setleft \reonode \in \reonodeset_2 \setbar \coloring_2(\reonode) = \colsymflow \setright \mbox{ and}
						\\	\reofnodeset_1 \cap \reonodeset_2 = \reofnodeset_2 \cap \reonodeset_1 \mbox{ and } \coloringtableindex = \tuple{\coloringtableindex_1 , \coloringtableindex_2}
						\end{array} \right. \right\}
			\\	=	& \pcomment{Because, by the Cartesian product, $[\coloringtableindex_1 \in \coloringtableindexset_1 \mbox{ and } \coloringtableindex_2 \in \coloringtableindexset_2]$ iff $\tuple{\coloringtableindex_1 , \coloringtableindex_2} \in \coloringtableindexset_1 \times \coloringtableindexset_2$}
			\\		& \left\{ \tuple{\coloringtableindex , \reofnodeset_1 \cup \reofnodeset_2 , \caconstraint_1 \wedge \caconstraint_2 , \tuple{\coloringtableindex_1' , \coloringtableindex_2'}} \left| \begin{array}{l}
							\ccoloring_1 = \tuple{\coloring_1 , \caconstraint_1} \in \ccoloringtablemap_1(\coloringtableindex_1) \mbox{ and } \coloringtableindex_1' = \cnextfun_1(\coloringtableindex_1 , \ccoloring_1)
							\mbox{ and } \reofnodeset_1 = \setleft \reonode \in \reonodeset_1 \setbar \coloring_1(\reonode) = \colsymflow \setright \mbox{ and}
						\\	\ccoloring_2 = \tuple{\coloring_2 , \caconstraint_2} \in \ccoloringtablemap_2(\coloringtableindex_2) \mbox{ and } \coloringtableindex_2' = \cnextfun_2(\coloringtableindex_2 , \ccoloring_2)
							\mbox{ and } \reofnodeset_2 = \setleft \reonode \in \reonodeset_2 \setbar \coloring_2(\reonode) = \colsymflow \setright \mbox{ and}
						\\	\reofnodeset_1 \cap \reonodeset_2 = \reofnodeset_2 \cap \reonodeset_1 \mbox{ and } \coloringtableindex = \tuple{\coloringtableindex_1 , \coloringtableindex_2} \in \coloringtableindexset_1 \times \coloringtableindexset_2
						\end{array} \right. \right\}
			\\	=	& \pcomment{Because, by the definition of $\reofnodeset_1$ and $\reofnodeset_2$ in Lemma \ref{lemma:dist:toca}, $[\reofnodeset_1 \cap \reonodeset_2 = \reofnodeset_2 \cap \reonodeset_1]$ iff $[\setleft \reonode \in \reonodeset_1 \cap \reonodeset_2 \setbar \coloring_1(\reonode) = \colsymflow \setright = \setleft \reonode \in \reonodeset_1 \cap \reonodeset_2 \setbar \coloring_2(\reonode) = \colsymflow \setright]$, and because, as $\coloring_1$ and $\coloring_2$ are 2-colorings, $[\setleft \reonode \in \reonodeset_1 \cap \reonodeset_2 \setbar \coloring_1(\reonode) = \colsymflow \setright = \setleft \reonode \in \reonodeset_1 \cap \reonodeset_2 \setbar \coloring_2(\reonode) = \colsymflow \setright]$ iff $[\coloring_1(\reonode) = \coloring_2(\reonode) \mbox{ for all } \reonode \in \reonodeset_1 \cap \reonodeset_2]$}
			\\		& \left\{ \tuple{\coloringtableindex , \reofnodeset_1 \cup \reofnodeset_2 , \caconstraint_1 \wedge \caconstraint_2 , \tuple{\coloringtableindex_1' , \coloringtableindex_2'}} \left| \begin{array}{l}
							\ccoloring_1 = \tuple{\coloring_1 , \caconstraint_1} \in \ccoloringtablemap_1(\coloringtableindex_1) \mbox{ and } \coloringtableindex_1' = \cnextfun_1(\coloringtableindex_1 , \ccoloring_1)
							\mbox{ and } \reofnodeset_1 = \setleft \reonode \in \reonodeset_1 \setbar \coloring_1(\reonode) = \colsymflow \setright \mbox{ and}
						\\	\ccoloring_2 = \tuple{\coloring_2 , \caconstraint_2} \in \ccoloringtablemap_2(\coloringtableindex_2) \mbox{ and } \coloringtableindex_2' = \cnextfun_2(\coloringtableindex_2 , \ccoloring_2)
							\mbox{ and } \reofnodeset_2 = \setleft \reonode \in \reonodeset_2 \setbar \coloring_2(\reonode) = \colsymflow \setright \mbox{ and}
						\\	{}[\coloring_1(\reonode) = \coloring_2(\reonode) \mbox{ for all } \reonode \in \reonodeset_1 \cap \reonodeset_2] \mbox{ and } \coloringtableindex = \tuple{\coloringtableindex_1 , \coloringtableindex_2} \in \coloringtableindexset_1 \times \coloringtableindexset_2
						\end{array} \right. \right\}
			\\	=	& \pcomment{Because, by Definition \ref{def:ccoloringtablecomp} of $\ccoloringtablecomp$, $[\ccoloring_1 = \tuple{\coloring_1 , \caconstraint_1} \in \ccoloringtablemap_1(\coloringtableindex_1) \mbox{ and } \ccoloring_2 = \tuple{\coloring_2 , \caconstraint_2} \in \ccoloringtablemap_2(\coloringtableindex_2) \mbox{ and } \coloring_1(\reonode) = \coloring_2(\reonode) \mbox{ for all } \reonode \in \reonodeset_1 \cap \reonodeset_2]$ iff $[\ccoloring_1 \ccoloringcomp \ccoloring_2 \in \ccoloringtablemap_1(\coloringtableindex_1) \ccoloringtablecomp \ccoloringtablemap_2(\coloringtableindex_2)]$}
			\\		& \left\{ \tuple{\coloringtableindex , \reofnodeset_1 \cup \reofnodeset_2 , \caconstraint_1 \wedge \caconstraint_2 , \tuple{\coloringtableindex_1' , \coloringtableindex_2'}} \left| \begin{array}{l}
							\ccoloring_1 = \tuple{\coloring_1 , \caconstraint_1} \mbox{ and } \coloringtableindex_1' = \cnextfun_1(\coloringtableindex_1 , \ccoloring_1)
							\mbox{ and } \reofnodeset_1 = \setleft \reonode \in \reonodeset_1 \setbar \coloring_1(\reonode) = \colsymflow \setright \mbox{ and}
						\\	\ccoloring_2 = \tuple{\coloring_2 , \caconstraint_2} \mbox{ and } \coloringtableindex_2' = \cnextfun_2(\coloringtableindex_2 , \ccoloring_2)
							\mbox{ and } \reofnodeset_2 = \setleft \reonode \in \reonodeset_2 \setbar \coloring_2(\reonode) = \colsymflow \setright \mbox{ and}
						\\	\ccoloring_1 \ccoloringcomp \ccoloring_2 \in \ccoloringtablemap_1(\coloringtableindex_1) \ccoloringtablecomp \ccoloringtablemap_2(\coloringtableindex_2) \mbox{ and } \coloringtableindex = \tuple{\coloringtableindex_1 , \coloringtableindex_2} \in \coloringtableindexset_1 \times \coloringtableindexset_2
						\end{array} \right. \right\}
			\\	=	& \pcomment{By Definition \ref{def:ccoloringtablemapcomp} of $\ccoloringtablemapcomp$ (informally on page \pageref{def:ccoloringtablemapcomp:info})}
			\\		& \left\{ \tuple{\coloringtableindex , \reofnodeset_1 \cup \reofnodeset_2 , \caconstraint_1 \wedge \caconstraint_2 , \tuple{\coloringtableindex_1' , \coloringtableindex_2'}} \left| \begin{array}{l}
							\ccoloring_1 = \tuple{\coloring_1 , \caconstraint_1} \mbox{ and } \coloringtableindex_1' = \cnextfun_1(\coloringtableindex_1 , \ccoloring_1)
							\mbox{ and } \reofnodeset_1 = \setleft \reonode \in \reonodeset_1 \setbar \coloring_1(\reonode) = \colsymflow \setright \mbox{ and}
						\\	\ccoloring_2 = \tuple{\coloring_2 , \caconstraint_2} \mbox{ and } \coloringtableindex_2' = \cnextfun_2(\coloringtableindex_2 , \ccoloring_2)
							\mbox{ and } \reofnodeset_2 = \setleft \reonode \in \reonodeset_2 \setbar \coloring_2(\reonode) = \colsymflow \setright \mbox{ and}
						\\	\ccoloring_1 \ccoloringcomp \ccoloring_2 \in (\ccoloringtablemap_1 \ccoloringtablemapcomp \ccoloringtablemap_2)(\coloringtableindex) \mbox{ and } \coloringtableindex = \tuple{\coloringtableindex_1 , \coloringtableindex_2} \in \coloringtableindexset_1 \times \coloringtableindexset_2
						\end{array} \right. \right\}
			\\	=	& \pcomment{Because, by Definition \ref{def:cnextfuncomp} of $\cnextfuncomp$ (informally on page \pageref{def:cnextfuncomp:info}), $[\coloringtableindex = \tuple{\coloringtableindex_1 , \coloringtableindex_2} \in \coloringtableindexset_1 \times \coloringtableindexset_2$ and $\ccoloring_1 \ccoloringcomp \ccoloring_2 \in (\ccoloringtablemap_1 \ccoloringtablemapcomp \ccoloringtablemap_2)(\coloringtableindex)$ and $\coloringtableindex_1' = \cnextfun_1(\coloringtableindex_1 , \ccoloring_1)$ and $\coloringtableindex_2' = \cnextfun_2(\coloringtableindex_2 , \ccoloring_2)]$ iff $[\tuple{\coloringtableindex_1', \coloringtableindex_2'} = (\cnextfun_1 \cnextfuncomp \cnextfun_2)(\coloringtableindex , \ccoloring_1 \ccoloringcomp \ccoloring_2)]$}
			\\		& \left\{ \tuple{\coloringtableindex , \reofnodeset_1 \cup \reofnodeset_2 , \caconstraint_1 \wedge \caconstraint_2 , \tuple{\coloringtableindex_1' , \coloringtableindex_2'}} \left| \begin{array}{l}
							\ccoloring_1 = \tuple{\coloring_1 , \caconstraint_1} \mbox{ and } \reofnodeset_1 = \setleft \reonode \in \reonodeset_1 \setbar \coloring_1(\reonode) = \colsymflow \setright \mbox{ and}
						\\	\ccoloring_2 = \tuple{\coloring_2 , \caconstraint_2} \mbox{ and } \reofnodeset_2 = \setleft \reonode \in \reonodeset_2 \setbar \coloring_2(\reonode) = \colsymflow \setright \mbox{ and}
						\\	\ccoloring_1 \ccoloringcomp \ccoloring_2 \in (\ccoloringtablemap_1 \ccoloringtablemapcomp \ccoloringtablemap_2)(\coloringtableindex) \mbox{ and } \coloringtableindex \in \coloringtableindexset_1 \times \coloringtableindexset_2 \mbox{ and } \tuple{\coloringtableindex_1', \coloringtableindex_2'} = (\cnextfun_1 \cnextfuncomp \cnextfun_2)(\coloringtableindex , \ccoloring_1 \ccoloringcomp \ccoloring_2)
						\end{array} \right. \right\}
			\\	=	& \pcomment{By introducing $\reofnodeset = \reofnodeset_1 \cup \reofnodeset_2$, and by applying $\tuple{\coloringtableindex_1', \coloringtableindex_2'} = (\cnextfun_1 \cnextfuncomp \cnextfun_2)(\coloringtableindex , \ccoloring_1 \ccoloringcomp \ccoloring_2)$}
			\\		& \left\{ \tuple{\coloringtableindex , \reofnodeset , \caconstraint_1 \wedge \caconstraint_2 , (\cnextfun_1 \cnextfuncomp \cnextfun_2)(\coloringtableindex , \ccoloring_1 \ccoloringcomp \ccoloring_2)} \left| \begin{array}{l}
							\ccoloring_1 = \tuple{\coloring_1 , \caconstraint_1} \mbox{ and } \ccoloring_2 = \tuple{\coloring_2 , \caconstraint_2} \mbox{ and}
						\\	\reofnodeset = \setleft \reonode \in \reonodeset_1 \cup \reonodeset_2 \setbar \coloring_1(\reonode) = \colsymflow \mbox{ or } \coloring_2(\reonode) = \colsymflow \setright \mbox{ and}
						\\	\ccoloring_1 \ccoloringcomp \ccoloring_2 \in (\ccoloringtablemap_1 \ccoloringtablemapcomp \ccoloringtablemap_2)(\coloringtableindex) \mbox{ and } \coloringtableindex \in \coloringtableindexset_1 \times \coloringtableindexset_2
						\end{array} \right. \right\}
			\\	=	& \pcomment{Because, by Definition \ref{def:coloringcomp} of $\coloringcomp$, $[\coloring_1(\reonode) = \colsymflow \mbox{ or } \coloring_2(\reonode) = \colsymflow]$ iff $[(\coloring_1 \coloringcomp \coloring_2)(\reonode) = \colsymflow]$, and by applying $\tuple{\coloring , \caconstraint} = \ccoloring = \ccoloring_1 \ccoloringcomp \ccoloring_2 = \tuple{\coloring_1 \coloringcomp \coloring_2 , \caconstraint_1 \wedge \caconstraint_2}$}
			\\		& \setleft \tuple{\coloringtableindex , \reofnodeset , \caconstraint , (\cnextfun_1 \cnextfuncomp \cnextfun_2)(\coloringtableindex , \ccoloring)} \setbar \reofnodeset = \setleft \reonode \in \reonodeset_1 \cup \reonodeset_2 \setbar \coloring(\reonode) = \colsymflow \setright \mbox{ and } \ccoloring = \tuple{\coloring , \caconstraint} \in (\ccoloringtablemap_1 \ccoloringtablemapcomp \ccoloringtablemap_2)(\coloringtableindex) \mbox{ and } \coloringtableindex \in \coloringtableindexset_1 \times \coloringtableindexset_2 \setright
			\\	=	& \pcomment{By the definition of $\catransrel'$ in Lemma \ref{lemma:dist:toca}}
			\\		& \catransrel'
			\end{array}
		$}}
		\caption{Proof: $\catransrel = \catransrel'$.}
		\label{fig:proof:lemma:dist:toca}
	\end{figure}
\end{lemma}

Although we consider only coloring models with two colors in this paper, one can apply Definition \ref{def:toca} of $\toca$ also to 3-colored $\cnextfuninit$-connectors. In fact, Lemma \ref{lemma:corr:bisim:toca} (bi-simulation) would still hold! Essentially, this means that 3-colored $\cnextfuninit$-connectors do not have a higher degree of expressiveness than coloring models with two colors. In contrast, Lemma \ref{lemma:dist:toca} (compositionality) does \emph{not} hold if we consider 3-colored $\cnextfuninit$-connectors. More precisely, the fourth|counted from top to bottom|equality in Figure \ref{fig:proof:lemma:dist:toca} (``Because, by the definition of $\reofnodeset_1$ and $\reofnodeset_2$...'') becomes invalid if we consider coloring models with three colors. This means that, although coloring models with two and three colors have the same degree of expressiveness, \emph{they compose differently}: paradoxically, the addition of a third color restricts, as intended, the number of compatible colorings. This allows us, for instance, to describe compositional context-sensitive connectors with three colors (considered impossible with two colors).

\newcommand{\coloringfromtransition}	{\mbox{\upshape\texttt{col}}}

\newcommand{\tocainv}	{\frac{1}{\mathbb{L}}}

%
\section{From \texorpdfstring{$\ca$}{alpha}-Connectors to \texorpdfstring{$\cnextfuninit$}{epsilon}-Connectors}
\label{sect:catocc}

In this section, we demonstrate a correspondence between $\cnextfuninit$-connectors and $\ca$-connectors in the direction opposite to the previous section's: from the latter to the former. Our approach, however, resembles our approach in Section \ref{sect:cctoca}: we present a unary operator, denoted by $\tocainv$, which takes as argument an $\ca$-connector and produces an equivalent|i.e., bi-similar|$\cnextfuninit$-connector. We call our process of transforming an $\ca$-connector to an $\nextfun$-connector the \emph{$\tocainv$-transformation} and define the $\tocainv$-operator for \emph{any} $\ca$-connector. It follows that the class of connectors that we can model as $\ca$-connector includes those that we can model as $\cnextfuninit$-connector. Since the previous section gave us a similar result in the opposite direction, we conclude that $\cnextfuninit$-connectors and $\ca$-connectors have the same degree of expressiveness.%

The $\tocainv$ operator works as follows; suppose we wish to transform an $\ca$-connector $\connca = \tuple{\connstruct, \ca}$ over $[\reonodeset , \caconstraintset]$. Whereas the connector structure $\connstruct$ does not incur any change (because $\tocainv$ alters only the behavioral model), from the CA $\ca$, the $\tocainv$-operator derives an initialized constraint next function: for each transition $\tuple{\castate , \reofnodeset , \caconstraint , \castate'}$ in the transition relation of $\ca$, $\tocainv$ includes a mapping from state $\castate$ and a constraint coloring $\ccoloring = \tuple{\coloring , \caconstraint}$ to state $\castate'$, where $\coloring$ assigns the flow color to all and only nodes in $\reofnodeset$.%
\begin{definition}
	[$\coloringfromtransition$]
	\label{def:coloringfromtransition}
	Let $\reonodeset , \reofnodeset \subseteq \reonodeuniverse$. Then:
	\begin{center}$
		\coloringfromtransition(\reonodeset , \reofnodeset) = \left\{ \reonode \mapsto \col \; \left| \; \reonode \in \reonodeset \mbox{ and } \col = \left( \begin{array}{@{}l@{\enspace}l@{}}
					\colsymflow		& \mbox{if } n \in \reofnodeset
				\\	\colsymnoflow	& \mbox{otherwise}
				\end{array} \right) \right. \right\}
	$\end{center}
\end{definition}
\begin{definition}
	[$\tocainv$ for $\ca$-connectors]
	\label{def:tocainv}
	Let $\connca = \tuple{\connstruct , \ca}$ be an $\ca$-connector over $[\reonodeset , \caconstraintset]$ with $\ca = \tuple{\castateset , \catransrel , \castate_0}$ a CA over $[\reonodeset , \caconstraintset]$. The $\tocainv$-transformation of $\connca$, denoted by $\tocainv(\connca)$, is defined as:
	\begin{center}$
		\begin{array}{@{}r@{\;}l@{}}
			\multicolumn{2}{c}{\tocainv(\connca) = \tuple{\connstruct , \tocainv(\ca)}}
		\\	\vspace{-.5em}
		\\	\mbox{with:}
		&	\tocainv(\ca) = \tuple{\cnextfun , \castate_0}
		\\	\mbox{and:}
		&	\cnextfun = \setleft \tuple{\castate , \tuple{\coloringfromtransition(\reonodeset , \reofnodeset) , \caconstraint}} \mapsto \castate' \setbar \tuple{\castate , \reofnodeset , \caconstraint , \castate'} \in \catransrel \setright
		\end{array}
	$\end{center}
\end{definition}
\noindent The following proposition states that the application of $\tocainv$ to an $\ca$-connector yields an $\cnextfuninit$-connector.%
\begin{proposition}
	\label{prop:wf:tocainv}
	Let $\connca = \tuple{\connstruct , \ca}$ be an $\ca$-connector over $[\reonodeset , \caconstraintset]$. Then, $\tocainv(\connca)$ is an $\cnextfuninit$-connector over $[\reonodeset , \ccoloringtablemap]$ with $\ccoloringtablemap$ defined over $[\reonodeset , \caconstraintset , \castateset]$ as:
	\begin{center}
		$\ccoloringtablemap = \setleft \castate \mapsto \ccoloringtable \setbar \castate \in \castateset \mbox{ and } \ccoloringtable = \setleft \tuple{\coloringfromtransition(\reonodeset , \reofnodeset) , \caconstraint} \setbar \tuple{\castate , \reofnodeset , \caconstraint , \castate'} \in \catransrel \setright \setright$
	\end{center}

	\barproof{%
		Let $\connca = \tuple{\connstruct , \ca}$ with $\ca = \tuple{\castateset , \catransrel , \castate_0}$ a CA over $[\reonodeset , \caconstraintset]$. Then, by Definition \ref{def:tocainv}, $\tocainv(\connca) = \tuple{\connstruct , \tocainv(\ca)} = \tuple{\cnextfun , \castate_0}$. By Definition \ref{def:connccolinit}, we must show that $\cnextfun$ is a constraint next function over $\ccoloringtablemap$. First, by Definition \ref{def:coloringfromtransition}, all colorings that occur in the domain of $\cnextfun$ have $\reonodeset$ as their domain. Next, by Definition \ref{def:tocainv}, all indexes that occur in the domain and co-domain of $\cnextfun$ are states that appear in elements of the transition relation $\catransrel$; therefore, by Definition \ref{def:ca}, all indexes come from $\castateset$. Similarly, by Definition \ref{def:tocainv}, all data constraints that occur in the domain of $\cnextfun$ also appear in elements of $\catransrel$, hence come from $\caconstraintset$. Finally, by Definition \ref{def:cnextfun}, we must show that $[\tuple{\coloringtableindex , \ccoloring} \mapsto \coloringtableindex'] \in \cnextfun$ iff $\ccoloring \in \ccoloringtablemap(\coloringtableindex)$. This follows straightforwardly from Definition \ref{def:tocainv} and the definition of $\ccoloringtablemap$ in this proposition.
	}
	\vspace{.25em}
\end{proposition}

\subsection{Inverse}
\label{sect:catocc:inv}

Having defined $\tocainv$, we proceed by proving that it forms the \emph{inverse} of $\toca$ (as already hinted at by its symbol) and vice versa. We do this before stating the correctness of $\tocainv$ and its distributivity over composition, because the proofs of these lemmas become significantly easier (and shorter) once we know that $\tocainv$ inverts $\toca$. The following two lemmas state the inversion properties in both directions.%
\begin{lemma}
	\label{lemma:inv:toca}
	Let $\connccolinit$ be an $\cnextfuninit$-connector. Then, $\tocainv(\toca(\connccolinit)) = \connccolinit$.
	
	\vspace{.5em}
	\barproof{%
		Let $\connccolinit = \tuple{\connstruct , \cnextfuninit}$ be defined over $[\reonodeset , \ccoloringtablemap]$ with $\ccoloringtablemap$ a constraint CTM over $[\reonodeset , \caconstraintset , \coloringtableindexset]$. By Definitions \ref{def:connccolinit}, \ref{def:toca}, and \ref{def:tocainv}, we must show that $\tocainv(\toca(\cnextfuninit)) = \cnextfuninit$. This follows from Figure \ref{fig:proof:lemma:inv:toca}.
	}
	\vspace{.25em}
	
	\begin{figure}[t]
		\centering
		\fbox{\scalebox{.75}{$
			\begin{array}{@{}c@{\;}l@{}}
					& \tocainv(\toca(\cnextfuninit))
			\\	=	& \pcomment{By Definition \ref{def:toca} of $\toca$, and because $\cnextfuninit$ is defined over $\ccoloringtablemap$, which is defined over $[\reonodeset , \caconstraintset , \coloringtableindexset]$}
			\\		& \tocainv(\tuple{\coloringtableindexset , \catransrel , \coloringtableindex_0}) \mbox{ with } \catransrel \mbox{ as in Definition \ref{def:toca}}
			\\	=	& \pcomment{By Definition \ref{def:tocainv} of $\tocainv$, and because $\toca(\cnextfuninit)$ is an $\ca$-connector over $[\reonodeset , \caconstraintset]$ by Proposition \ref{prop:wf:toca}}
			\\		& \tuple{\setleft \tuple{\coloringtableindex , \tuple{\coloringfromtransition(\reonodeset , \reofnodeset) , \caconstraint}} \mapsto \coloringtableindex' \setbar \tuple{\coloringtableindex , \reofnodeset , \caconstraint , \coloringtableindex'} \in \catransrel \setright , \coloringtableindex_0} \mbox{ with } \catransrel \mbox{ as in Definition \ref{def:toca}}
			\\	=	& \pcomment{Because $\tuple{\coloringtableindex , \reofnodeset , \caconstraint , \coloringtableindex'} \in \catransrel$ iff $[\coloringtableindex \in \coloringtableindexset \mbox{ and } \ccoloring = \tuple{\coloring , \caconstraint} \in \ccoloringtablemap(\castate) \mbox{ and }$ \\ $\reofnodeset = \setleft \reonode \in \reonodeset \setbar \coloring(\reonode) = \colsymflow \setright \mbox{ and } \coloringtableindex' = \cnextfun(\coloringtableindex , \ccoloring)]$ by Definition \ref{def:toca} of $\toca$}
			\\		& \tuple{\setleft \tuple{\coloringtableindex , \tuple{\coloringfromtransition(\reonodeset , \reofnodeset) , \caconstraint}} \mapsto \cnextfun(\coloringtableindex , \ccoloring) \setbar \coloringtableindex \in \coloringtableindexset \mbox{ and } \ccoloring = \tuple{\coloring , \caconstraint} \in \ccoloringtablemap(\coloringtableindex) \mbox{ and } \reofnodeset = \setleft \reonode \in \reonodeset \setbar \coloring(\reonode) = \colsymflow \setright \setright , \coloringtableindex_0}
			\\	=	& \pcomment{Because $\coloring = \coloringfromtransition(\reonodeset , \reofnodeset) \mbox{ iff } \reofnodeset = \setleft \reonode \in \reonodeset \setbar \coloring(\reonode) = \colsymflow \setright$ by Definitions \ref{def:coloring} and \ref{def:coloringfromtransition}}
			\\		& \tuple{\setleft \tuple{\coloringtableindex , \tuple{\coloring , \caconstraint}} \mapsto \cnextfun(\coloringtableindex , \ccoloring) \setbar \coloringtableindex \in \coloringtableindexset \mbox{ and } \ccoloring = \tuple{\coloring , \caconstraint} \in \ccoloringtablemap(\coloringtableindex) \setright , \coloringtableindex_0}
			\\	=	& \pcomment{By Definitions \ref{def:cnextfun} and \ref{def:cnextfuninit}}
			\\		& \tuple{\cnextfun , \coloringtableindex_0} = \cnextfuninit
			\end{array}
		$}}
		\caption{Proof: $\tocainv(\toca(\cnextfuninit)) = \cnextfuninit$.}
		\label{fig:proof:lemma:inv:toca}
	\end{figure}
\end{lemma}

\begin{lemma}
	\label{lemma:inv:tocainv}
	Let $\connca$ be an $\ca$-connector. Then, $\toca(\tocainv(\connca)) = \connca$.
	
	\vspace{.5em}
	\barproof{%
		Let $\connca = \tuple{\connstruct , \ca}$ be defined over $[\reonodeset , \caconstraintset]$. By Definitions \ref{def:connca}, \ref{def:toca}, and \ref{def:tocainv}, we must show that $\toca(\tocainv(\ca)) = \ca$. This follows from Figure \ref{fig:proof:lemma:inv:tocainv}.
	}
	
	\begin{figure}[t]
		\centering
		\fbox{\scalebox{.75}{$
			\begin{array}{@{}c@{\;}l@{}}
					& \toca(\tocainv(\ca))
			\\	=	& \pcomment{By Definition \ref{def:tocainv} of $\tocainv$, and because $\ca = \tuple{\castateset , \catransrel , \castate_0}$ is a constraint automaton over $[\reonodeset , \caconstraintset]$ by the premise of Lemma \ref{lemma:inv:tocainv}}
			\\		& \toca(\tuple{\cnextfun , \castate_0}) \mbox{ with } \cnextfun = \setleft \tuple{\castate , \tuple{\coloringfromtransition(\reonodeset , \reofnodeset) , \caconstraint}} \mapsto \castate' \setbar \tuple{\castate , \reofnodeset , \caconstraint , \castate'} \in \catransrel \setright
			\\	=	& \pcomment{By Definition \ref{def:toca} of $\toca$, and because $\cnextfun$ is defined over $\ccoloringtablemap$ with \\ $\ccoloringtablemap = \setleft \castate \mapsto \ccoloringtable \setbar \castate \in \castateset \mbox{ and } \ccoloringtable = \setleft \tuple{\coloringfromtransition(\reonodeset , \reofnodeset) , \caconstraint} \setbar \tuple{\castate , \reofnodeset , \caconstraint , \castate'} \in \catransrel \setright \setright$ by Proposition 2}
			\\		& \tuple{\castateset , \catransrel' , \castate_0} \mbox{ with:}
			\\		& \;\bullet\enspace \catransrel' = \setleft \tuple{\castate , \reofnodeset , \caconstraint , \cnextfun(\castate , \ccoloring)} \setbar \castate \in \castateset \mbox{ and } \ccoloring = \tuple{\coloring , \caconstraint} \in \ccoloringtablemap(\castate) \mbox{ and } \reofnodeset = \setleft \reonode \in \reonodeset \setbar \coloring(\reonode) = \colsymflow \setright \setright
			\\		& \;\bullet\enspace \cnextfun = \setleft \tuple{\castate , \tuple{\coloringfromtransition(\reonodeset , \reofnodeset) , \caconstraint}} \mapsto \castate' \setbar \tuple{\castate , \reofnodeset , \caconstraint , \castate'} \in \catransrel \setright
			\\		& \;\bullet\enspace \ccoloringtablemap = \setleft \castate \mapsto \ccoloringtable \setbar \castate \in \castateset \mbox{ and } \ccoloringtable = \setleft \tuple{\coloringfromtransition(\reonodeset , \reofnodeset) , \caconstraint} \setbar \tuple{\castate , \reofnodeset , \caconstraint , \castate'} \in \catransrel \setright \setright
			\\	=	& \pcomment{By introducing $\castate' = \cnextfun(\castate , \ccoloring)$ in $\catransrel'$}
			\\		& \tuple{\castateset , \catransrel' , \castate_0} \mbox{ with:}
			\\		& \;\bullet\enspace \catransrel' = \setleft \tuple{\castate , \reofnodeset , \caconstraint , \castate'} \setbar \castate \in \castateset \mbox{ and } \ccoloring = \tuple{\coloring , \caconstraint} \in \ccoloringtablemap(\castate) \mbox{ and } \reofnodeset = \setleft \reonode \in \reonodeset \setbar \coloring(\reonode) = \colsymflow \setright \mbox{ and } \castate' = \cnextfun(\castate , \ccoloring) \setright
			\\		& \;\bullet\enspace \cnextfun = \setleft \tuple{\castate , \tuple{\coloringfromtransition(\reonodeset , \reofnodeset) , \caconstraint}} \mapsto \castate' \setbar \tuple{\castate , \reofnodeset , \caconstraint , \castate'} \in \catransrel \setright
			\\		& \;\bullet\enspace \ccoloringtablemap = \setleft \castate \mapsto \ccoloringtable \setbar \castate \in \castateset \mbox{ and } \ccoloringtable = \setleft \tuple{\coloringfromtransition(\reonodeset , \reofnodeset) , \caconstraint} \setbar \tuple{\castate , \reofnodeset , \caconstraint , \castate'} \in \catransrel \setright \setright
			\\	=	& \pcomment{Because, by the definition of $\ccoloringtablemap$, $[\ccoloring = \tuple{\coloring , \caconstraint} \in \ccoloringtablemap(\castate)]$ iff $[\tuple{\castate , \reofnodeset' , \caconstraint , \castate''} \in \catransrel \mbox{ and } \coloring = \coloringfromtransition(\reonodeset , \reofnodeset')]$}
			\\		& \tuple{\castateset , \catransrel' , \castate_0} \mbox{ with:} 
			\\		& \;\bullet\enspace \catransrel' = \setleft \tuple{\castate , \reofnodeset , \caconstraint , \castate'} \setbar \castate \in \castateset \mbox{ and } \tuple{\castate , \reofnodeset' , \caconstraint , \castate''} \in \catransrel \mbox{ and } \coloring = \coloringfromtransition(\reonodeset , \reofnodeset') \mbox{ and } \reofnodeset = \setleft \reonode \in \reonodeset \setbar \coloring(\reonode) = \colsymflow \setright \mbox{ and } \castate' = \cnextfun(\castate , \tuple{\coloring , \caconstraint}) \setright
			\\		& \;\bullet\enspace \cnextfun = \setleft \tuple{\castate , \tuple{\coloringfromtransition(\reonodeset , \reofnodeset) , \caconstraint}} \mapsto \castate' \setbar \tuple{\castate , \reofnodeset , \caconstraint , \castate'} \in \catransrel \setright
			\\		& \pcomment{By applying $\coloring = \coloringfromtransition(\reonodeset , \reofnodeset')$}
			\\		& \;\bullet\enspace \catransrel' = \setleft \tuple{\castate , \reofnodeset , \caconstraint , \castate'} \setbar \castate \in \castateset \mbox{ and } \tuple{\castate , \reofnodeset' , \caconstraint , \castate''} \in \catransrel \mbox{ and } \reofnodeset = \setleft \reonode \in \reonodeset \setbar (\coloringfromtransition(\reonodeset , \reofnodeset'))(\reonode) = \colsymflow \setright \mbox{ and } \castate' = \cnextfun(\castate , \tuple{\coloringfromtransition(\reonodeset , \reofnodeset') , \caconstraint}) \setright
			\\		& \;\bullet\enspace \cnextfun = \setleft \tuple{\castate , \tuple{\coloringfromtransition(\reonodeset , \reofnodeset) , \caconstraint}} \mapsto \castate' \setbar \tuple{\castate , \reofnodeset , \caconstraint , \castate'} \in \catransrel \setright
			\\		& \pcomment{Because, by Definition \ref{def:coloringfromtransition} of $\coloringfromtransition$, $\setleft \reonode \in \reonodeset \setbar (\coloringfromtransition(\reonodeset , \reofnodeset'))(\reonode) = \colsymflow \setright = \reofnodeset'$}
			\\		& \tuple{\castateset , \catransrel' , \castate_0} \mbox{ with:} 
			\\		& \;\bullet\enspace \catransrel' = \setleft \tuple{\castate , \reofnodeset , \caconstraint , \castate'} \setbar \castate \in \castateset \mbox{ and } \tuple{\castate , \reofnodeset' , \caconstraint , \castate''} \in \catransrel \mbox{ and } \reofnodeset = \reofnodeset' \mbox{ and } \castate' = \cnextfun(\castate , \tuple{\coloringfromtransition(\reonodeset , \reofnodeset') , \caconstraint}) \setright
			\\		& \;\bullet\enspace \cnextfun = \setleft \tuple{\castate , \tuple{\coloringfromtransition(\reonodeset , \reofnodeset) , \caconstraint}} \mapsto \castate' \setbar \tuple{\castate , \reofnodeset , \caconstraint , \castate'} \in \catransrel \setright
			\\		& \pcomment{By applying $\reofnodeset = \reofnodeset'$}
						\\		& \tuple{\castateset , \catransrel' , \castate_0} \mbox{ with:} 
			\\		& \;\bullet\enspace \catransrel' = \setleft \tuple{\castate , \reofnodeset' , \caconstraint , \castate'} \setbar \castate \in \castateset \mbox{ and } \tuple{\castate , \reofnodeset' , \caconstraint , \castate''} \in \catransrel \mbox{ and } \castate' = \cnextfun(\castate , \tuple{\coloringfromtransition(\reonodeset , \reofnodeset') , \caconstraint}) \setright
			\\		& \;\bullet\enspace \cnextfun = \setleft \tuple{\castate , \tuple{\coloringfromtransition(\reonodeset , \reofnodeset) , \caconstraint}} \mapsto \castate' \setbar \tuple{\castate , \reofnodeset , \caconstraint , \castate'} \in \catransrel \setright
			\\	=	& \pcomment{Because, by the definition of $\cnextfun$, $\castate' = \cnextfun(\castate , \tuple{\coloringfromtransition(\reonodeset , \reofnodeset') , \caconstraint})$ iff $\tuple{\castate , \reofnodeset' , \caconstraint , \castate'} \in \catransrel$}
			\\		& \tuple{\castateset , \catransrel' , \castate_0} \mbox{ with } \catransrel' = \setleft \tuple{\castate , \reofnodeset' , \caconstraint , \castate'} \setbar \castate \in \castateset \mbox{ and } \tuple{\castate , \reofnodeset' , \caconstraint , \castate''} \in \catransrel \mbox{ and } \tuple{\castate , \reofnodeset' , \caconstraint , \castate'} \in \catransrel \setright = \catransrel
			\\	=	& \pcomment{By the definition of $\ca$}
			\\		& \ca
			\end{array}
		$}}
		\caption{Proof: $\toca(\tocainv(\ca)) = \ca$.}
		\label{fig:proof:lemma:inv:tocainv}
	\end{figure}
\end{lemma}

\subsection{Correctness and Distributivity of \texorpdfstring{$\tocainv$}{1/L}}
\label{sect:catocc:corrdist}

As mentioned previously, knowing that $\toca(\tocainv(\connca)) = \connca$ (with $\connca$ an $\ca$-connector) greatly simplifies our correctness and distributivity proofs. We start with the former. Lemma \ref{lemma:corr:bisim:tocainv}, which appears below, states the bi-similarity between $\connca$ and its $\tocainv$-transformation $\tocainv(\connca)$. In addition to the inversion lemma, in our proof, we apply the bi-similarity lemma of $\toca$.%
\begin{lemma}
	\label{lemma:corr:bisim:tocainv}
	Let $\connca$ be an $\ca$-connector. Then, $\connca \tocabisim \tocainv(\connca)$.
	
	\vspace{.5em}
	\barproof{
		Let $\connccolinit = \tocainv(\connca)$. Then, $\connca \tocabisim \tocainv(\connca) \mbox{ iff } \toca(\tocainv(\connca)) \tocabisim \tocainv(\connca) \mbox{ iff } \toca(\connccolinit) \tocabisim \connccolinit$ by Lemma \ref{lemma:inv:tocainv}. The latter, $\toca(\connccolinit) \tocabisim \connccolinit$, follows from Lemma \ref{lemma:corr:bisim:toca}.
	}
	\vspace{.25em}
\end{lemma}
\noindent Finally, Lemma \ref{lemma:dist:tocainv} states the distributivity of $\tocainv$ over composition: informally, this means that it does not matter whether we first compose $\ca$-connectors $\connca_1$ and $\connca_2$ and then apply $\tocainv$ to the composition or first apply $\tocainv$ to $\connca_1$ and $\connca_2$ and then compose the transformations; the resulting $\cnextfuninit$-connectors equal each other. Our proof, similar to that of the previous lemma, relies on the inversion lemmas.
\begin{lemma}
	\label{lemma:dist:tocainv}
	Let $\connca_1$ and $\connca_2$ be $\ca$-connectors. Then, $\tocainv(\connca_1) \connccolinitcomp \tocainv(\connca_2) = \tocainv(\connca_1 \conncacomp \connca_2)$.
	
	\vspace{.5em}
	\barproof{
		Follows from Figure \ref{fig:proof:lemma:dist:tocainv}.
	}
	
	\begin{figure}[t]
		\centering
		\fbox{\scalebox{.75}{$
			\begin{array}{@{}c@{\;}l@{}}
					& \tocainv(\connca_1) \connccolinitcomp \tocainv(\connca_2)
			\\	=	& \pcomment{By the inversion of $\toca$ by $\tocainv$ in Lemma \ref{lemma:inv:toca}}
			\\		& \tocainv(\toca(\tocainv(\connca_1) \connccolinitcomp \tocainv(\connca_2)))
			\\	=	& \pcomment{By the distributivity of $\toca$ over composition in Lemma \ref{lemma:dist:toca}}
			\\		& \tocainv(\toca(\tocainv(\connca_1)) \conncacomp \toca(\tocainv(\connca_2)))
			\\	=	& \pcomment{By the inversion of $\tocainv$ by $\toca$ in Lemma \ref{lemma:inv:tocainv}}
			\\		& \tocainv(\connca_1 \conncacomp \connca_2)
			\end{array}
		$}}
		\caption{Proof: $\tocainv(\connca_1) \connccolinitcomp \tocainv(\connca_2) = \tocainv(\connca_1 \conncacomp \connca_2)$.}
		\label{fig:proof:lemma:dist:tocainv}
	\end{figure}
\end{lemma}

%
\section{Application}
\label{sect:appl}

In this section, we sketch an application of the results presented above: the integration of verification and animation of context-sensitive connectors in Vereofy \cite{baier09}, a model checking tool for $\alpha$-connectors that operates on constraint automata.%
\footnote{%
	Vereofy is freely available on-line at: \url{http://www.vereofy.de}.
} Broadly, this application consists of two parts: model checking connectors built from context-sensitive constituents and generating animated counterexamples.

\paragraph{Verification of \texorpdfstring{$\cnextfuninit$}{epsilon}-connectors}

Vereofy operates on constraint automata and, therefore, many consider it unable to verify context-sensitive connectors. We mend this deficiency as follows. First, we note that recent research established that one can transform coloring models with three colors, known for their ability to properly capture context-sensitivity, to corresponding coloring models with two colors \cite{jongmans11}. Essentially, this means that $\cnextfuninit$-connectors as defined in this paper|i.e., featuring only two colors|can serve as faithful models of context-sensitive circuits. Consequently, the results in Section \ref{sect:cctoca} enable the verification of such connectors with Vereofy: using the $\toca$-transformation, we transform context-sensitive $\cnextfuninit$-connectors to context-sensitive $\ca$-connectors, whose CA we subsequently can analyze with Vereofy. In this application, the distributivity of $\toca$ over composition in Lemma \ref{lemma:dist:toca} plays an important role (as already outlined in Section \ref{sect:cctoca:dist}): it facilitates (i) the one-time-application of $\toca$ to the context-sensitive $\cnextfuninit$-connectors of Reo's primitives after which (ii) we can use Vereofy's built-in functionality for CA composition to construct the complex automata that we wish to inspect. Examples appear in \cite{jongmans11}. The distributivity lemmas work also in the opposite direction: if future studies indicate that composition of coloring models costs less than composition of CA, we may extend Vereofy with a module to automatically (1) transform CA of primitives to coloring models with $\tocainv$, (2) compose the resulting coloring models, and (3) transform the resulting composition back to a CA with $\toca$. (To truly gain in performance, however, the costs of transforming forth and back should not exceed the benefits of composing coloring models instead of CA.)

\paragraph{Animation of \texorpdfstring{$\ca$}{alpha}-connectors}

Vereofy facilitates the generation and inspection of counterexamples, an important feature that distinguishes it from mCRL2 (another tool sometimes used for model checking Reo circuits \cite{kokash10}).%
\footnote{%
	mCRL2 is freely available on-line at: \url{http://www.mcrl2.org}.
} When using Vereofy in conjunction with the Eclipse Coordination Tools (Reo's standard distribution),%
\footnote{%
	The Eclipse Coordination Tools are freely available on-line at \url{http://reo.project.cwi.nl}.
} it can in \emph{some} cases display counterexamples as connector animations. These animated counterexamples comprise a graphical model of a connector (similar to Figure \ref{fig:connstruct:prim}) through which data items visually flow for each computation step of a faulty behavior. Although this approach greatly enhances the ease with which users can analyze counterexamples, the opportunity to actually provide these animations depends on the availability of a coloring model of the connector under investigation (in addition to the constraint automaton that Vereofy's verification algorithm operates on). Moreover, the standalone version of Vereofy, a command-line tool, does not facilitate the animation of counterexamples at all. The results in Section \ref{sect:catocc}, however, enable animated counterexamples for \emph{any} CA: in the case of unavailability of a coloring model, Vereofy can simply generate such a model with the $\tocainv$-transformation.

%
\section{Concluding Remarks}
\label{sect:conc}

\paragraph{Related work}

Closest to the work in this paper seems an informal discussion in \cite{clarke07,costa10} on the equivalence of coloring models and constraint automata. These cited references, however, do not support their claims with formal evidence, nor do they provide an algorithm, operation, or function to actually transform connector models back and forth. More generally, we know of only a few other correspondences between different semantic models of Reo connectors, the oldest concerning CA and coalgebraic models: the set of runs of an $\ca$-connector $\connca$ corresponds to the set of \emph{timed data streams} induced by the coalgebraic model of the same circuit that $\connca$ models (Definition 3.6 in \cite{baier06}). Reo's semantics in terms of the \emph{unifying theories of programming} \cite{meng09} appears closely related to the coalgebraic semantics as well, but we do not know of any formal claims or proofs. Two other correspondences concern \emph{tile models} \cite{arbab09} and coloring models: Arbab et al. prove in Theorems 1 and 3 of \cite{arbab09} that a semantic model based on \emph{tiles}, which resemble colorings, yields behavioral formalisms equal to coloring models with two or three colors (depending on the tile definitions).

\paragraph{Conclusion and future work} 

We showed that, once extended with data constraints, coloring models with two colors and constraint automata have the same degree of expressiveness by defining two operators that transform data-aware coloring models to equivalent CA and vice versa. Moreover, these operators distribute over composition, a desirable property especially from a practical point of view. Though primarily a theoretical contribution, we illustrated how our results can broaden the applicability of Reo's tools. With respect to future work, we would like to implement the transformation operators and the sketched extension to Vereofy. Another application worth investigation comprises the development of an implementation of Reo based on transforming the behavioral models of connectors back and forth. Finally, we would like to study correspondences between other semantic models (e.g., guarded automata \cite{bonsangue09} and intentional automata \cite{costa10}).

\paragraph{Acknowledgments}

We would like to thank the reviewers and the members of the ICE 2011 discussion forum \texttt{gege}, \texttt{wind}, \texttt{wolf} and \texttt{xyz} for their valuable comments.

\bibliographystyle{eptcs}
\bibliography{generic}

\begin{thebibliography}{10}
\providecommand{\bibitemdeclare}[2]{}
\providecommand{\urlprefix}{Available at }
\providecommand{\url}[1]{\texttt{#1}}
\providecommand{\href}[2]{\texttt{#2}}
\providecommand{\urlalt}[2]{\href{#1}{#2}}
\providecommand{\doi}[1]{doi:\urlalt{http://dx.doi.org/#1}{#1}}
\providecommand{\bibinfo}[2]{#2}

\bibitemdeclare{article}{arbab04}
\bibitem{arbab04}
\bibinfo{author}{Farhad Arbab} (\bibinfo{year}{2004}):
  \emph{\bibinfo{title}{Reo: {A} channel-based coordination model for component
  composition}}.
\newblock {\sl \bibinfo{journal}{Mathematical Structures in Computer Science}}
  \bibinfo{volume}{14}(\bibinfo{number}{3}), pp. \bibinfo{pages}{329--366},
  \doi{10.1017/S0960129504004153}.

\bibitemdeclare{article}{arbab05}
\bibitem{arbab05}
\bibinfo{author}{Farhad Arbab} (\bibinfo{year}{2005}):
  \emph{\bibinfo{title}{Abstract behavior types: {A} foundation model for
  components and their composition}}.
\newblock {\sl \bibinfo{journal}{Science of Computer Programming}}
  \bibinfo{volume}{55}(\bibinfo{number}{1--3}), pp. \bibinfo{pages}{3--52},
  \doi{10.1016/j.scico.2004.05.010}.

\bibitemdeclare{incollection}{arbab09}
\bibitem{arbab09}
\bibinfo{author}{Farhad Arbab}, \bibinfo{author}{Roberto Bruni},
  \bibinfo{author}{Dave Clarke}, \bibinfo{author}{Ivan Lanese} \&
  \bibinfo{author}{Ugo Montanari} (\bibinfo{year}{2009}):
  \emph{\bibinfo{title}{Tiles for {R}eo}}.
\newblock In \bibinfo{editor}{Andrea Corradini} \& \bibinfo{editor}{Ugo
  Montanari}, eds.: {\sl \bibinfo{booktitle}{Recent Trends in Algebraic
  Development Techniques}}, {\sl \bibinfo{series}{LNCS}}
  \bibinfo{volume}{5486}, \bibinfo{publisher}{Springer}, pp.
  \bibinfo{pages}{37--55}, \doi{10.1007/978-3-642-03429-9\_4}.

\bibitemdeclare{incollection}{arbab03}
\bibitem{arbab03}
\bibinfo{author}{Farhad Arbab} \& \bibinfo{author}{Jan Rutten}
  (\bibinfo{year}{2003}): \emph{\bibinfo{title}{A coinductive calculus of
  component connectors}}.
\newblock In \bibinfo{editor}{Marin Wirsing}, \bibinfo{editor}{Dirk Pattinson}
  \& \bibinfo{editor}{Rolf Hennicker}, eds.: {\sl \bibinfo{booktitle}{Recent
  Trends in Algebraic Development Techniques}}, {\sl \bibinfo{series}{LNCS}}
  \bibinfo{volume}{2755}, \bibinfo{publisher}{Springer}, pp.
  \bibinfo{pages}{34--55}, \doi{10.1007/978-3-540-40020-2\_2}.

\bibitemdeclare{incollection}{baier09}
\bibitem{baier09}
\bibinfo{author}{Christel Baier}, \bibinfo{author}{Tobias Blechmann},
  \bibinfo{author}{Joachim Klein} \& \bibinfo{author}{Sascha Kl\"{u}ppelholz}
  (\bibinfo{year}{2009}): \emph{\bibinfo{title}{Formal verification for
  components and connectors}}.
\newblock In \bibinfo{editor}{Frank de~Boer}, \bibinfo{editor}{Marcello
  Bonsangue} \& \bibinfo{editor}{Eric Madelaine}, eds.: {\sl
  \bibinfo{booktitle}{Formal Methods for Components and Objects}}, {\sl
  \bibinfo{series}{LNCS}} \bibinfo{volume}{5751},
  \bibinfo{publisher}{Springer}, pp. \bibinfo{pages}{82--101},
  \doi{10.1007/978-3-642-04167-9\_5}.

\bibitemdeclare{article}{baier06}
\bibitem{baier06}
\bibinfo{author}{Christel Baier}, \bibinfo{author}{Marjan Sirjani},
  \bibinfo{author}{Farhad Arbab} \& \bibinfo{author}{Jan Rutten}
  (\bibinfo{year}{2006}): \emph{\bibinfo{title}{Modeling component connectors
  in {R}eo by constraint automata}}.
\newblock {\sl \bibinfo{journal}{Science of Computer Programming}}
  \bibinfo{volume}{61}(\bibinfo{number}{2}), pp. \bibinfo{pages}{75--113},
  \doi{10.1016/j.scico.2005.10.008}.

\bibitemdeclare{incollection}{bonsangue09}
\bibitem{bonsangue09}
\bibinfo{author}{Marcello Bonsangue}, \bibinfo{author}{Dave Clarke} \&
  \bibinfo{author}{Alexandra Silva} (\bibinfo{year}{2009}):
  \emph{\bibinfo{title}{Automata for context-dependent connectors}}.
\newblock In \bibinfo{editor}{John Field} \& \bibinfo{editor}{Vasco
  Vasconcelos}, eds.: {\sl \bibinfo{booktitle}{Coordination Models and
  Languages}}, {\sl \bibinfo{series}{LNCS}} \bibinfo{volume}{5521},
  \bibinfo{publisher}{Springer}, pp. \bibinfo{pages}{184--203},
  \doi{10.1007/978-3-642-02053-7\_10}.

\bibitemdeclare{article}{clarke07}
\bibitem{clarke07}
\bibinfo{author}{Dave Clarke}, \bibinfo{author}{David Costa} \&
  \bibinfo{author}{Farhad Arbab} (\bibinfo{year}{2007}):
  \emph{\bibinfo{title}{Connector colouring {I}: {S}ynchronisation and context
  dependency}}.
\newblock {\sl \bibinfo{journal}{Science of Computer Programming}}
  \bibinfo{volume}{66}(\bibinfo{number}{3}), pp. \bibinfo{pages}{205--225},
  \doi{10.1016/j.scico.2007.01.009}.

\bibitemdeclare{phdthesis}{costa10}
\bibitem{costa10}
\bibinfo{author}{David Costa} (\bibinfo{year}{2010}):
  \emph{\bibinfo{title}{Formal Models for Component Connectors}}.
\newblock Ph.D. thesis, \bibinfo{school}{Vrije Universiteit Amsterdam}.

\bibitemdeclare{incollection}{jongmans11}
\bibitem{jongmans11}
\bibinfo{author}{Sung-Shik Jongmans}, \bibinfo{author}{Christian Krause} \&
  \bibinfo{author}{Farhad Arbab} (\bibinfo{year}{2011}):
  \emph{\bibinfo{title}{Encoding context-sensitivity in {R}eo into
  non-context-sensitive semantic models}}.
\newblock In \bibinfo{editor}{Wolfgang de~Meuter} \& \bibinfo{editor}{Catalin
  Roman}, eds.: {\sl \bibinfo{booktitle}{Proceedings of the 13th International
  Conference on Coordination Models and Languages}}, {\sl
  \bibinfo{series}{LNCS}} \bibinfo{volume}{6721},
  \bibinfo{publisher}{Springer}, pp. \bibinfo{pages}{31--48},
  \doi{10.1007/978-3-642-21464-6\_3}.

\bibitemdeclare{inproceedings}{koehler09}
\bibitem{koehler09}
\bibinfo{author}{Christian Koehler} \& \bibinfo{author}{Dave Clarke}
  (\bibinfo{year}{2009}): \emph{\bibinfo{title}{Decomposing port automata}}.
\newblock In: {\sl \bibinfo{booktitle}{Proceedings of the 2009 ACM Symposium on
  Applied Computing}}, pp. \bibinfo{pages}{1369--1373},
  \doi{10.1145/1529282.1529587}.

\bibitemdeclare{incollection}{kokash10}
\bibitem{kokash10}
\bibinfo{author}{Natallia Kokash}, \bibinfo{author}{Christian Krause} \&
  \bibinfo{author}{Erik de~Vink} (\bibinfo{year}{2010}):
  \emph{\bibinfo{title}{Verification of context-dependent channel-based service
  models}}.
\newblock In \bibinfo{editor}{Frank de~Boer}, \bibinfo{editor}{Marcello
  Bonsangue}, \bibinfo{editor}{Stefan Hallerstede} \& \bibinfo{editor}{Michael
  Leuschel}, eds.: {\sl \bibinfo{booktitle}{Formal Methods for Components and
  Objects}}, {\sl \bibinfo{series}{LNCS}} \bibinfo{volume}{6286},
  \bibinfo{publisher}{Springer}, pp. \bibinfo{pages}{21--40},
  \doi{10.1007/978-3-642-17071-3\_2}.

\bibitemdeclare{article}{meng09}
\bibitem{meng09}
\bibinfo{author}{Sun Meng} \& \bibinfo{author}{Farhad Arbab}
  (\bibinfo{year}{2009}): \emph{\bibinfo{title}{Connectors as designs}}.
\newblock {\sl \bibinfo{journal}{ENTCS}} \bibinfo{volume}{255}, pp.
  \bibinfo{pages}{119--135}, \doi{10.1016/j.entcs.2009.10.028}.

\bibitemdeclare{phdthesis}{proenca11}
\bibitem{proenca11}
\bibinfo{author}{Jos\'{e} Proe\c{c}a} (\bibinfo{year}{2011}):
  \emph{\bibinfo{title}{Synchronous Coordination of Distributed Components}}.
\newblock Ph.D. thesis, \bibinfo{school}{Universiteit Leiden}.

\end{thebibliography}

\appendix

%
\newpage
\section{Appendix: Composition Operators}
\label{appx}

In this appendix, we give the formal definitions of the composition operators whose definition we gave only informally in Section \ref{sect:constraintcc}. More specifically, we give the definitions of the composition operators for constraint CTMs, (initialized) constraint next functions, and $\cnextfuninit$-connectors. The definitions of the composition operators for constraint colorings and constraint coloring tables appear in Section \ref{sect:constraintcc}. As mentioned in that section, we obtain the operators that we define below by replacing $\coloringtablemap_1$, $\coloringtablemap_2$, $\nextfun_1$, $\nextfun_2$, $\nextfuninit_1$, and $\nextfuninit_2$ in Definitions \ref{def:coloringtablemapcomp}--\ref{def:conncolinitcomp} by their $\boldsymbol{font}$ versions $\ccoloringtablemap_1$, $\ccoloringtablemap_2$, $\cnextfun_1$, $\cnextfun_2$, $\cnextfuninit_1$, and $\cnextfuninit_2$.
\begin{definition}
	[Composition of constraint CTMs]
	\label{def:ccoloringtablemapcomp}
	Let $\ccoloringtablemap_1$ and $\ccoloringtablemap_2$ be constraint CTMs over $[\reonodeset_1 , \caconstraintset_1 , \coloringtableindexset_1]$ and $[\reonodeset_2 , \caconstraintset_2 , \coloringtableindexset_2]$. Their composition, denoted by $\ccoloringtablemap_1 \ccoloringtablemapcomp \ccoloringtablemap_2$, is a constraint CTM over $[\reonodeset_1 \cup \reonodeset_2 , \caconstraintset_1 \wedge \caconstraintset_2 , \coloringtableindexset_1 \times \coloringtableindexset_2]^{\ref{footnote:wedge}}$ defined as:
	\begin{center}$
		\ccoloringtablemap_1 \ccoloringtablemapcomp \ccoloringtablemap_2 = \setleft \tuple{\coloringtableindex_1 , \coloringtableindex_2} \mapsto \ccoloringtablemap_1(\coloringtableindex_1) \ccoloringtablecomp \ccoloringtablemap_2(\coloringtableindex_2) \setbar \coloringtableindex_1 \in \coloringtableindexset_1 \mbox{ and } \coloringtableindex_2 \in \coloringtableindexset_2 \setright
 	$\end{center}
\end{definition}
\begin{definition}
	[Composition of constraint next functions]
	\label{def:cnextfuncomp}
	Let $\cnextfun_1$ and $\cnextfun_2$ be constraint next functions over $\ccoloringtablemap_1$ and $\ccoloringtablemap_2$ with $\ccoloringtablemap_1$ and $\ccoloringtablemap_2$ constraint CTMs over $[\reonodeset_1 , \caconstraintset_1 , \coloringtableindexset_1]$ and $[\reonodeset_2 , \caconstraintset_2 , \coloringtableindexset_2]$. Their composition, denoted by $\cnextfun_1 \cnextfuncomp \cnextfun_2$, is a next function over $\ccoloringtablemap_1 \ccoloringtablemapcomp \ccoloringtablemap_2$ defined as:
	\begin{center}$
		\begin{array}{@{}l@{\;}c@{\;}l@{}}
			\cnextfun_1 \cnextfuncomp \cnextfun_2 & = & \left\{ \left. \begin{array}{@{}c@{\;}}
				\tuple{\coloringtableindex_1 , \coloringtableindex_2} , \ccoloring_1 \ccoloringcomp \ccoloring_2 
			\\	\rotatebox[origin=c]{270}{$\mapsto$}
			\\	\tuple{\cnextfun_1(\coloringtableindex_1 , \ccoloring_1) , \cnextfun_2(\coloringtableindex_2 , \ccoloring_2)}
			\end{array} \right| \begin{array}{@{\;}c@{}}
				\tuple{\coloringtableindex_1 , \coloringtableindex_2} \in \coloringtableindexset_1 \times \coloringtableindexset_2
			\\	\mbox{and}
			\\	\ccoloring_1 \ccoloringcomp \ccoloring_2 \in (\ccoloringtablemap_1 \ccoloringtablemapcomp \ccoloringtablemap_2)(\tuple{\coloringtableindex_1 , \coloringtableindex_2})
			\end{array} \right\}
		\end{array}
	$\end{center}
\end{definition}
\begin{definition}
	[Composition of initialized constraint next functions]
	\label{def:cnextfuninitcomp}
	Let $\cnextfuninit_1 = \tuple{\cnextfun_1 , \coloringtableindex_0^1}$ and $\cnextfuninit_2 = \tuple{\cnextfun_2 , \coloringtableindex_0^2}$ be initialized constraint next functions over $\ccoloringtablemap_1$ and $\ccoloringtablemap_2$. Their composition, denoted by $\cnextfuninit_1 \cnextfuninitcomp \cnextfuninit_2$, is an initialized next function over $[\ccoloringtablemap_1 \ccoloringtablemapcomp \ccoloringtablemap_2]$ defined as:
	\begin{center}$
		\cnextfuninit_1 \cnextfuninitcomp \cnextfuninit_2 = \tuple{\cnextfun_1 \cnextfuncomp \cnextfun_2 , \tuple{\coloringtableindex_0^1 , \coloringtableindex_0^2}}
	$\end{center}
\end{definition}
\begin{definition}
	[Composition of $\cnextfuninit$-connectors]
	\label{def:connccolinitcomp}
	Let $\connccolinit_1 = \tuple{\connstruct_1 , \cnextfuninit_1}$ and $\connccolinit_2 = \tuple{\connstruct_2 , \cnextfuninit_2}$ be $\cnextfuninit$-connectors over $[\reonodeset_1 , \ccoloringtablemap_1]$ and $[\reonodeset_2 , \ccoloringtablemap_2]$ such that $\connstruct_1 \connstructcomp \connstruct_2$ is defined. Their composition, denoted by $\connccolinit_1 \connccolinitcomp \connccolinit_2$, is an $\cnextfuninit$-connector over $[\reonodeset_1 \cup \reonodeset_2 , \ccoloringtablemap_1 \ccoloringtablemapcomp \ccoloringtablemap_2]$ defined as:
	\begin{center}$
		\connccolinit_1 \connccolinitcomp \connccolinit_2 = \tuple{\connstruct_1 \connstructcomp \connstruct_2 , \cnextfuninit_1 \cnextfuninitcomp \cnextfuninit_2}
	$\end{center}
\end{definition}

\end{document}